\documentclass[prb,superscriptaddress,showpacs,floatfix,tightenlines,10pt,twocolumn]{revtex4}
\usepackage{amssymb}
\usepackage{amsmath}
\usepackage{graphicx}

\begin{document}

\title{Orbital order in bilayer graphene at filling factor $\nu =-1 $}
\author{R. C\^{o}t\'{e} }
\affiliation{D\'{e}partement de physique, Universit\'{e} de Sherbrooke, Sherbrooke, Qu%
\'{e}bec, J1K 2R1, Canada}
\author{Jules Lambert}
\affiliation{D\'{e}partement de physique, Universit\'{e} de Sherbrooke, Sherbrooke, Qu%
\'{e}bec, J1K 2R1, Canada}
\author{Yafis Barlas}
\affiliation{National High Magnetic Field Laboratory and Department of Physics, The
Florida State University, Florida 32306, USA}
\author{A. H. MacDonald}
\affiliation{Department of Physics, The University of Texas at Austin, Austin, Texas
78712, USA}
\keywords{graphene}
\pacs{73.21.-b,73.22.Gk,78.70.Gq}

\begin{abstract}
In a graphene bilayer with Bernal stacking both $n=0$ and $n=1$ orbital
Landau levels have zero kinetic energy. An electronic state in the $N=0$
Landau level consequently has three quantum numbers in addition to its
guiding center label: its spin, its valley index $K$ or $K^{\prime }$, and
an orbital quantum number $n=0,1.$ The two-dimensional electron gas (2DEG)
in the bilayer supports a wide variety of broken-symmetry states in which
the pseudospins associated these three quantum numbers order in a manner
that is dependent on both filling factor $\nu $ and the electric potential
difference between the layers. In this paper, we study the case of $\nu =-1$
in an external field strong enough to freeze electronic spins. We show that
an electric potential difference between layers drives a series of
transitions, starting from interlayer-coherent states (ICS) at small
potentials and leading to orbitally coherent states (OCS) that are polarized
in a single layer. Orbital pseudospins carry electric dipoles with
orientations that are ordered in the OCS and have Dzyaloshinskii-Moriya
interactions that can lead to spiral instabilities. We show that the
microwave absorption spectra of ICSs, OCSs, and the mixed states that occur
at intermediate potentials are sharply distinct.
\end{abstract}

\date{\today }
\maketitle

\section{INTRODUCTION}

Semiconductor double-quantum-well systems in a quantizing magnetic field
develop spontaneous inter-layer coherence when the wells are brought into
close proximity.\cite{coherencerefs} Spontaneous coherence leads to a
variety of fascinating transport effects including counterflow superfluidity
and anomalous interlayer tunneling, and to unusual charged excitations such
as merons. A convenient way to describe these ground states is to use a
pseudospin language in which the \emph{which layer} degree-of-freedom is
mapped to a $S=1/2$ pseudospin. In this language, the ground state of a
bilayer at total filling factor $\nu =1$ is an easy-plane pseudospin
ferromagnet. At higher filling factors, still more exotic states occur , for
example states in which the pseudospin orientation varies in space and a
charge-density-wave is formed.\cite{doublestripes}

Interest has recently been growing in the strong-magnetic-field ordered
states of graphene bilayers. Single layer graphene\cite{graphenereviews} is
a two-dimensional honeycomb lattice network of carbon atoms. Bilayer graphene%
\cite{bilayerexpt,mccann,koshino} consists of two graphene layers separated
by a fraction of a nanometer. In the normal Bernal stacking structure, one
of the two honeycomb sublattice sites in each layer has a near-neighbor in
the other layer, and one does not. This arrangement produces\cite%
{bilayerexpt,mccann} a set of Landau levels with energies $E_{N}=\pm \hslash
\omega _{c}^{\ast }\sqrt{\left\vert N\right\vert \left( \left\vert
N\right\vert +1\right) }$ where $\omega _{c}^{\ast }$ is the effective
cyclotron frequency and $N=0,\pm 1,\pm 2,...$. All Landau levels except $N=0$
are four-fold degenerate; electronic states are specified by $N$,
valley-index ($K$ or $K^{\prime }$) and spin-index, in addition to the usual
label used to specify guiding center states within a Landau level.

The $N=0$ Landau level has an additional two-valued quantum
degree-of-freedom because states with both $n=0$ and $n=1$ Landau-level
character have zero kinetic energy. Most of the new physics discussed in
this paper is related to the property\cite{BarlasPRL} that electric dipoles
can be constructed by forming wavefunctions with coherence between $n=0$ and 
$n=1$ components. A second peculiarity of the $N=0$ state is that
wavefunctions associated with the $K$ valley are localized in one layer,
while wavefunctions associated with the $K^{\prime }$ valley are localized
in the opposite layer. Layer and valley indices are thus equivalent. It is
convenient to use pseudospins to represent both layer (or equivalently
valley) and the Landau-level orbital character degrees of freedom. The
wavefunction for an electron in the $N=0$ Landau level is therefore the
direct product of a standard guiding center factor and three spinors that
capture its dependence on spin, layer, and orbital-Landau-index ($n$)
character. We refer to the final spinor as the orbital spinor, and to the
set of eight Landau levels with zero kinetic energy as the bilayer graphene
octet.\cite{BarlasPRL} In neutral graphene the octet is half-filled at all
magnetic field strengths.

The presence of the octet in bilayers is revealed experimentally by a jump
in the quantized Hall conductivity \cite{bilayerexpt} from $-4(e^{2}/h)$ to $%
4(e^{2}/h)$ when the charge density is tuned across neutrality in moderately
disordered samples. In a recent paper\cite{BarlasPRL} some of us predicted
that quantum Hall effects would appear at all integer filling factors
between $\nu =-4$ and $\nu =4$ in samples of quality sufficient\cite%
{NomuraPRL} to make interactions dominant relative to unintended disorder.
Electron-electron interactions acting alone are expected to lift the
degeneracy of the bilayer octet and induce gaps at the Fermi level by
producing a set of spontaneously broken symmetry states with spin, valley
and orbital pseudospin polarizations. The octet degeneracy lifting is
expected\cite{BarlasPRL} to follow a set of Hund's rules in which spin
polarization is maximized first, then layer polarization to the greatest
extent possible, and finally orbital polarization to the extent allowed by
the first two rules. Hall plateaus at all integer filling factors
intermediate between $\nu =-4$ and $\nu =4$ have indeed now been discovered
in experimental studies of suspended bilayer graphene samples and bilayer
graphene on SiO$_{2}$/Si substrates,\cite{octetexperiment,octetnewsandviews}
opening up the opportunity to study a rich and still relatively unexplored%
\cite{yafis2,shizuya1} family of novel broken symmetry states. The odd
filling factor cases are expected to be most interesting because all three
pseudospins are expected to be polarized. The present paper focuses on the
physics associated with the competition between layer and orbital
pseudospins at Landau levels $\nu =-1$ and $\nu =3$ at field strengths
sufficient to produce maximal spin polarization and reduce the importance of
Landau-level mixing. In this limit a negative filling factor $\nu $ is
equivalent to a positive filling factor $\nu +4$ since the two states differ
only through the presence in the latter case of inert filled majority spin
Landau levels.

In a previous paper\cite{yafis2} we studied the quantum Hall states which
occur at $\nu =-3$ and $\nu=1$ in the same strong field regime, emphasizing
the key role played by the potential energy difference between graphene
layers which we refer to here as the bias potential $\Delta_{B}$. The $%
\nu=-3 $ ground state at zero bias is an inter-layer coherent state with
orbital index $n=0$ that supports counterflow superfluidity. One
particularly interesting property of this state is that the superfluid
density, the coefficient that relates the counterflow supercurrent to the
spatial gradient of interlayer phase, vanishes. Correspondingly, the state's
Goldstone mode dispersion is quadratic in wavector $q$, in contrast to the
linear dispersion found in coherent semiconductor bilayers and in standard
superfluids. We also found that the uniform ground state has a
long-wavelength instability at any non-zero potential-difference bias $%
\Delta _{B}<\Delta _{B}^{\left( c\right) }$ where $\Delta _{B}^{\left(
c\right) }$ is the critical bias at which all $N=0$ charge is tranferred to
a single layer. In Ref. \onlinecite{yafis2}, we argued that the instability
is probably towards a state in which the direction of the inter-layer
pseudospin varies in space. For larger bias $\Delta _{B}>\Delta _{B}^{\left(
c\right) }$ the ground state is uninteresting; the charge is completely in
one layer (or valley) and in the orbital state $n=0.$ The orbital
pseudospinwave mode corresponding to transitions between the $n=0$ and $n=1$
orbital states is gapped at a frequency $\omega =\hslash \omega _{c}^{\ast
}\Delta _{B}/\gamma _{1}$ where $\gamma _{1}$ is the inter-layer tunneling
energy in the Bernal stacking. This mode, \textit{which is an intra-Landau
level excitation}, has a finite oscillator strength and will absorb\cite%
{bilayerCR} electromagnetic radiation. This behavior contrasts with the
standard Kohn's theorem\cite{Kohn} behavior in normal 2DEG's which implies
that only inter-Landau level excitations produce absorption.

Surprisingly the phase diagrams for $\nu =-1$ and $\nu=3$ states differ
qualitatively from the corresponding $\nu=-3$ and $\nu=1$ phase diagrams.
The source of the difference is a competition in the $\nu=-1$ case between
interaction and single-particle effects which are reinforcing in the $\nu=-3$
case. The end result is that the large $\Delta_{B}$ ground state at $\nu=-1$
places electrons in a coherent combination of $n=0$ and $n=1$ orbital
states, and that electric dipoles are consequently spontaneously present in
the ground state. This paper analyzes the dependence of bilayer properties
on $\Delta_{B}$ and explores some of the consequences of the unusual
orbitally ordered dipole state.

At small bias we find that the bilayer's $\nu=-1$ ground state is an
inter-layer coherent state, much like the corresponding $\nu=-3$ state
except that the coherence is between orbitals with $n=1$ character. This
state has a gapless pseudospin wave mode with linear dispersion, like
coherent semiconductor bilayers. The state also has a gapped orbital
pseudospin collective mode. Because the orbital spinor carries an electric
dipole, this mode has a finite oscillator strength and absorbs
electromagnetic radiation, again much like the $\nu=-3$ case. This mode
should be visible in a microwave spectoscopy experiment.

Inter-layer coherence decreases with bias until a new ground state is
reached that has both inter-layer and orbital coherences. In this mixed
state, the low-energy orbital and inter-layer pseudospin modes are both
gapped. Because the modes are coupled, both show up in the microwave
absorption spectra. The collective excitations are highly anisotropic in
this phase and we find that the intensity of the absorption depends strongly
on the orientation of the electric field of the incident microwaves.

The new physics of the $\nu =-1$ case emerges in its simplest form at still
stronger bias potentials. Both orbital levels in the bottom layer are then
completely filled while only one of the two top layer Landau levels is
filled. Spontaneous orbital coherence then develops in the top layer. This
spontaneous orbital coherence leads to a \textit{gapless} orbital pseudospin
mode. Some of the properties of this state have been studied independently
in a recent paper by Shizuya\cite{shizuya1}, who also pointed out that
orbital coherence is responsible for the existence of a finite density of
electrical dipoles with a net polarization. These dipoles collectively and
spontaneously point in some arbitrary direction in the $x-y$ plane. As
discussed by Shizuya\cite{shizuya1} their orientation can however be
controlled by an external electric field parallel to the plane of the
bilayer. In this paper, we use an effective pseudospin model to highlight
other interesting features of the orbitally-coherent state. In particular we
demonstrate the presence of a Dzyaloshinskii-Moriya (DM) interaction\cite{DM}
between orbital pseudospins and show that it leads to an anisotropic
softening of the orbital pseudospin mode at a finite wavevector. For strong
enough inter-layer bias, the DM induces an instability toward a pseudospin
spiral state. The orbital pseudospin mode in the high bias regime is gapless
and will lead, in the presence of disorder, to strong absorption of
electromagnetic waves at very small frequencies.

Our paper is organized in the following way. In Section II, we discuss the
non-interacting states of the graphene bilayer within a two-band low-energy
model. Here we introduce the aspect of the electronic structure that is
responsible for interaction and band effects which are competing at $\nu =-1$
and are reinforcing at $\nu =-3$. In Section III, we derive the Hamiltonian
of the graphene two-dimensional electron gas (2DEG) truncated to $N=0$
levels in the Hartree-Fock approximation. We use this Hamiltonian to derive
the equation of motion for the single-particle Green's function in Section
IV and to obtain the order parameters for the various phases which occur at $%
\nu =-1$. Section V\ describes the generalized random-phase approximation
(GRPA)\ (or equivalently the time-dependent Hartree-Fock approximation
(TDHFA)) that we use to derive the collective excitations. The phase diagram
for $\nu =-1$ as a function of bias is obtained in Section VI. We then study
the collective excitations of the inter-layer coherent phase in Section VII\
and those of the orbital coherent phase in Section VIII. Finally microwave
absorption in the different phases is studied in Section IX and we conclude
with a brief summary and some suggestions for future work in Section X.

\section{EFFECTIVE TWO-BAND HAMILTONIAN}

In a graphene bilayer with Bernal stacking, the two basis atoms of the top
layer are denoted by $A_{1}$ and $B_{1}$ and those of the bottom layer by $%
A_{2}$ and $B_{2}$ with atoms $A_{1}$ sitting directly above atoms $B_{2}$.
The band structure of the bilayer is calculated using a tight-binding model
with in-plane nearest-neighbor tunneling (with strength $\gamma _{0}=2.5$
eV) and $A_{1}-B_{2}$ tunneling (with strength $\gamma _{1}=0.4$ eV). (For a
review of bilayer graphene, see Ref. \onlinecite{reviewtheory}.) The
low-energy ($E<<\gamma _{1}$) excitations of this model for electrons in the
valleys $\xi \mathbf{K}$ with $\mathbf{K}=\frac{2\pi }{a_{0}}\left( -\frac{2%
}{3},0\right) $ and $\xi =\pm 1$ can be studied by using the effective
two-band model developped in Ref. \onlinecite{mccann}. Using the basis $%
\left\{ \left\vert A_{2}\right\rangle ,\left\vert B_{1}\right\rangle
\right\} $ for $H_{\mathbf{K}}^{0}$ and $\left\{ \left\vert
B_{1}\right\rangle ,\left\vert A_{2}\right\rangle \right\} $ for $H_{-%
\mathbf{K}}^{0}$, the effective two-band Hamiltonian derived in Ref. %
\onlinecite{mccann} is

\begin{equation}
H_{\xi \mathbf{K}}^{0}=\left( 
\begin{array}{cc}
-\xi \frac{\Delta _{B}}{2}+\xi \frac{\Delta _{B}}{\gamma _{1}}\frac{1}{%
2m^{\ast }}p_{-}p_{+} & \frac{1}{2m^{\ast }}p_{-}^{2} \\ 
\frac{1}{2m^{\ast }}p_{+}^{2} & \xi \frac{\Delta _{B}}{2}-\xi \frac{\Delta
_{B}}{\gamma _{1}}\frac{1}{2m^{\ast }}p_{+}p_{-}%
\end{array}%
\right) ,  \label{one}
\end{equation}%
where $p_{\pm }=p_{x}\pm ip_{y}$ and $\mathbf{p}=-i\hslash \nabla .$ In this
equation, $\Delta _{B}$ is the bias potential between the two layers, the
effective mass $m^{\ast }=2\hslash ^{2}\gamma _{1}/3\gamma
_{0}^{2}a_{0}^{2}=0.054m_{0}$ with $m_{0}$ the bare electronic mass, $a_{0}=%
\sqrt{3}c$ is the triangular lattice constant and $c=1.42$ \AA\ is the
distance between neighboring carbon atoms in the same plane. The kets $%
\left\vert A_{2}\right\rangle ,\left\vert B_{1}\right\rangle $ correspond to
the atomic sites in different layers that are not directly above one another.

The Hamiltonian of the 2DEG in a perpendicular magnetic field is obtained by
making the substitution $\mathbf{p}\rightarrow \mathbf{p}+e\mathbf{A}%
/c\equiv \mathbf{P}/\hslash $ (with $e>0$) in Eq. (\ref{one}). The vector
potential $\mathbf{A}$ is defined such that $\nabla \times \mathbf{A=B=}B%
\widehat{\mathbf{z}}.\mathbf{\ }$ In a magnetic field,%
\begin{equation}
H_{\xi \mathbf{K}}^{0}=\left( 
\begin{array}{cc}
-\xi \frac{1}{2}\Delta _{B}+\xi \beta \Delta _{B}\left( 1+a^{\dag }a\right)
& \hslash \omega _{c}^{\ast }a^{2} \\ 
\hslash \omega _{c}^{\ast }\left( a^{\dag }\right) ^{2} & \xi \frac{1}{2}%
\Delta _{B}-\xi \beta \Delta _{B}a^{\dag }a%
\end{array}%
\right) ,  \label{deux}
\end{equation}%
where we have defined the orbital ladder operators $a=\ell \left(
P_{x}-iP_{y}\right) /\sqrt{2}\hslash ,a^{\dag }=\ell \left(
P_{x}+iP_{y}\right) /\sqrt{2}\hslash $ with the magnetic length $\ell =\sqrt{%
\hslash c/eB}$ and the parameter $\beta =\hslash \omega _{c}^{\ast }/\gamma
_{1}=6.144\times 10^{-3}B$(Tesla). The effective cyclotron frequency $\omega
_{c}^{\ast }=eB/m^{\ast }c.$ At zero bias, the Landau levels have energies $%
E_{N}^{0}=\pm \xi \hslash \omega _{c}^{\ast }\sqrt{\left\vert N\right\vert
\left( \left\vert N\right\vert +1\right) }$ with $N=0,\pm 1,\pm 2,...$

In this paper, we study the phase diagram of the 2DEG in the $N=0$ Landau
level. While levels with $\left\vert N\right\vert >0$ are four-fold
degenerate (counting spin and valley quantum numbers), level $N=0$ has an
extra \textit{orbital} degeneracy due to the fact that Landau level orbitals 
$n=0$ and $n=1$ have zero kinetic energy. The states in $N=0$ are thus
member of an octet of Landau levels that are degenerate if we neglect the
Zeeman and bias potential energies. We assume that the Zeeman coupling is
strong enough to assure maximal spin-polarization, which allows this
degree-of-freedom to be neglected. The eigenfunctions and corresponding
energies for $N=0$ are then given by%
\begin{eqnarray}
\left( 
\begin{array}{c}
0 \\ 
h_{0,X}\left( \mathbf{r}\right)%
\end{array}%
\right) ,\;E_{\mathbf{K},n=0,X} &=&\frac{1}{2}\Delta _{B}, \\
\left( 
\begin{array}{c}
0 \\ 
h_{1,X}\left( \mathbf{r}\right)%
\end{array}%
\right) ,\;E_{\mathbf{K},n=1,X} &=&\frac{1}{2}\Delta _{B}-\beta \Delta _{B},
\end{eqnarray}%
for the $\mathbf{K}$ valley and by 
\begin{eqnarray}
\left( 
\begin{array}{c}
h_{0,X}\left( \mathbf{r}\right) \\ 
0%
\end{array}%
\right) ,\;E_{\mathbf{K}^{\prime },n=0,X} &=&-\frac{1}{2}\Delta _{B}, \\
\left( 
\begin{array}{c}
h_{1,X}\left( \mathbf{r}\right) \\ 
0%
\end{array}%
\right) ,\;E_{\mathbf{K}^{\prime },n=1,X} &=&-\frac{1}{2}\Delta _{B}+\beta
\Delta _{B}
\end{eqnarray}%
for the $\mathbf{K}^{\prime }$ valley, using this time the basis $\left\{
\left\vert A_{2}\right\rangle ,\left\vert B_{1}\right\rangle \right\} $ for
all states. It is quite clear from these equations that the valley $\mathbf{K%
}(\mathbf{K}^{\prime })$ eigenstates are localized in the top(bottom) layer.
For $N=0$, the layer index is thus equivalent to the valley index. (For $%
\left\vert N\right\vert >0$, the spinors have different orbital indices $n$
in different layers.) The functions $h_{n,X}\left( \mathbf{r}\right)
=e^{-iXy/\ell ^{2}}\varphi _{n}\left( x-X\right) /\sqrt{L_{y}}$ are the
Landau gauge ($\mathbf{A}=\left( 0,Bx,0\right) $) eigenstates of an electron
with guiding center $X$, and $\varphi _{n}\left( x\right) $ is the wave
function of a one-dimensionnal harmonic oscillator. Note that with our
choice of gauge, the action of the ladder operators on the states $\varphi
_{n}\left( x\right) $ is given by $a^{\dag }\varphi _{n}\left( x\right) =i%
\sqrt{n+1}\varphi _{n+1}\left( x\right) $ and $a\varphi _{n}\left( x\right)
=-i\sqrt{n}\varphi _{n-1}\left( x\right) .$

\begin{figure}[tbph]
\includegraphics[scale=1]{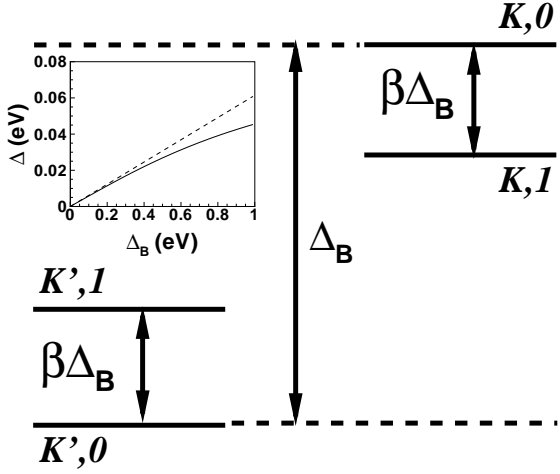}
\caption{ Non-interacting energy levels in Landau level $N=0.$ The inset
shows the behavior of the gap evaluated in the original (solid line) and
two-band (dashed line) models with bias.}
\label{niveaux}
\end{figure}

At finite bias, the parameter $\beta <<1$ lifts the degeneracy between the
two orbital states as we show in Fig. \ref{niveaux}. The splitting is
however very small. For positive bias, the $n=0$ orbital state in the
bottom(top) layer is lower(higher) in energy than the $n=1$ orbital state.
The orbital states $n=0,1$ form a two-level system in each valley and we
associate them with an \textit{orbital }pseudospin. Similarly, the two
states $\pm \mathbf{K}$ are associated with a \textit{valley} pseudospin. We
remark that the effective two-band model slightly overestimates the gap $%
\Delta =E_{n=1}-E_{n=0}$ between the $n=0$ and $n=1$ orbital states. The
inset in Fig. \ref{niveaux} shows the difference between the gap calculated
in the two-band model and in the original four-band system. In the region
where the DM interaction driven instability occurs, the difference between
the two gaps is however very small.

\section{HARTREE-FOCK HAMILTONIAN}

We now add the Coulomb interaction to the non-interacting Hamiltonian $%
H_{\xi \mathbf{K}}^{0}$. We assume that the magnetic field is strong enough
so that we can restrict the Hilbert space to the $N=0$ Landau level and
neglect Landau level mixing. We also assume the 2DEG to be fully spin
polarized (we comment on this later). We write the electron field operator
as 
\begin{eqnarray}
\Psi _{\mathbf{K}}\left( \mathbf{r}\right) &=&\sum_{X}\left( 
\begin{array}{c}
0 \\ 
h_{0,X}\left( \mathbf{r}\right)%
\end{array}%
\right) c_{K,X,0}  \label{field1} \\
&&+\sum_{X}\left( 
\begin{array}{c}
0 \\ 
h_{1,X}\left( \mathbf{r}\right)%
\end{array}%
\right) c_{K,X,1}  \notag
\end{eqnarray}%
and%
\begin{eqnarray}
\Psi _{\mathbf{K}^{\prime }}\left( \mathbf{r}\right) &=&\sum_{X}\left( 
\begin{array}{c}
h_{0,X}\left( \mathbf{r}\right) \\ 
0%
\end{array}%
\right) c_{K^{\prime },X,0}  \label{field2} \\
&&+\sum_{X}\left( 
\begin{array}{c}
h_{1,X}\left( \mathbf{r}\right) \\ 
0%
\end{array}%
\right) c_{K^{\prime },X,1},  \notag
\end{eqnarray}%
so that the Hartree-Fock Hamiltonian is given by (here and in the rest of
this paper, we use the convention that repeated indices are summed over)

\begin{gather}
H_{HF}=N_{\phi }E_{a,n}\rho _{n,n}^{a,a}\left( 0\right)  \label{hamil} \\
+N_{\phi }\overline{\sum_{\mathbf{q}}}H_{n_{1},n_{2},n_{3},n_{4}}^{a,b}%
\left( \mathbf{q}\right) \left\langle \rho _{n_{1},n_{2}}^{a,a}\left( -%
\mathbf{q}\right) \right\rangle \rho _{n_{3},n_{4}}^{b,b}\left( \mathbf{q}%
\right)  \notag \\
-N_{\phi }\sum_{\mathbf{q}}X_{n_{1},n_{4},n_{3},n_{2}}^{a,b}\left( \mathbf{q}%
\right) \left\langle \rho _{n_{1},n_{2}}^{a,b}\left( -\mathbf{q}\right)
\right\rangle \rho _{n_{3},n_{4}}^{b,a}\left( \mathbf{q}\right) ,  \notag
\end{gather}%
where $N_{\phi }$ is the Landau level degeneracy and all energies are
measured in units of $e^{2}/\kappa \ell $ where $\kappa $ is the effective
dielectric constant at the position of the graphene layers. The
single-particle energies $E_{a,n}$ include capacitive contributions and are
defined by 
\begin{equation}
E_{a,n}=\frac{1}{2}a\Delta _{B}-a\beta \Delta _{B}n+\left[ \frac{\widetilde{%
\nu }}{2}\frac{d}{\ell }-\widetilde{\nu }_{a}\frac{d}{\ell }\right] ,
\label{ener}
\end{equation}%
with $\widetilde{\nu }_{a}$ the number of filled levels in valley $a$, $%
\widetilde{\nu }=\nu +4$ the total number of filled levels, $a,b=\pm 1$ the
valley (or equivalently layer) index and $n=0,1$ for the two orbital state
indices. In deriving Eq. (\ref{hamil}), we have taken into account a
neutralizing positive background so that the $\mathbf{q}=0$ contribution is
absent in the Hartree term. This convention is indicated by the bar over the
summation. Note that for positive bias, the bottom layer ($\mathbf{K}%
^{\prime }$ valley) is at a lower potential than the top layer ($\mathbf{K}$
valley).

The density operators in Eq. (\ref{hamil}), are defined by

\begin{eqnarray}
\rho _{n_{1},n_{2}}^{a,b}\left( \mathbf{q}\right) &=&\frac{1}{N_{\phi }}%
\sum_{X_{1},X_{2}}e^{-\frac{i}{2}q_{x}\left( X_{1}+X_{2}\right) } \\
&&\times c_{a,X_{1},n_{1}}^{\dagger }c_{b,X_{2},n_{2}}\delta
_{X_{1},X_{2}+q_{y}\ell ^{2}},  \notag
\end{eqnarray}%
where $c_{a,X_{1},n_{1}}^{\dagger }$ creates an electron in state $\left(
a,X_{1},n_{1}\right) $ in the Landau gauge. The intralayer $\left(
H,X=H^{a,a},X^{a,a}\right) $ and inter-layer $\left( \widetilde{H},%
\widetilde{X}=H^{a\neq b},X^{a\neq b}\right) $ Hartree and Fock interactions
are given by 
\begin{eqnarray}
H_{n_{1},n_{2},n_{3},n_{4}}\left( \mathbf{q}\right) &=&\frac{1}{q\ell }%
K_{n_{1},n_{2}}\left( \mathbf{q}\right) K_{n_{3},n_{4}}\left( -\mathbf{q}%
\right) ,  \label{hart1} \\
X_{n_{1},n_{2},n_{3},n_{4}}\left( \mathbf{q}\right) &=&\int \frac{d\mathbf{p}%
\ell ^{2}}{2\pi }H_{n_{1},n_{2},n_{3},n_{4}}\left( \mathbf{p}\right) e^{i%
\mathbf{q}\times \mathbf{p}\ell ^{2}},  \label{fock1}
\end{eqnarray}%
and%
\begin{eqnarray}
\widetilde{H}_{n_{1},n_{2},n_{3},n_{4}}\left( \mathbf{q}\right)
&=&H_{n_{1},n_{2},n_{3},n_{4}}\left( \mathbf{q}\right) e^{-qd},
\label{hart2} \\
\widetilde{X}_{n_{1},n_{2},n_{3},n_{4}}\left( \mathbf{q}\right) &=&\int 
\frac{d\mathbf{p}\ell ^{2}}{2\pi }\widetilde{H}_{n_{1},n_{2},n_{3},n_{4}}%
\left( \mathbf{p}\right) e^{i\mathbf{q}\times \mathbf{p}\ell ^{2}},
\label{fock2}
\end{eqnarray}%
where $d=3.337$ \AA\ is the inter-layer separation in the Bernal stacking.
The form factors which appear here,%
\begin{eqnarray}
K_{0,0}\left( \mathbf{q}\right) &=&\exp \left( \frac{-q^{2}\ell ^{2}}{4}%
\right) ,  \label{K1} \\
K_{1,1}\left( \mathbf{q}\right) &=&\exp \left( \frac{-q^{2}\ell ^{2}}{4}%
\right) \left( 1-\frac{q^{2}\ell ^{2}}{2}\right) ,  \label{K2} \\
K_{1,0}\left( \mathbf{q}\right) &=&\left( \frac{\left( q_{y}+iq_{x}\right)
\ell }{\sqrt{2}}\right) \exp \left( \frac{-q^{2}\ell ^{2}}{4}\right) ,
\label{K3} \\
K_{0,1}\left( \mathbf{q}\right) &=&\left( \frac{\left( -q_{y}+iq_{x}\right)
\ell }{\sqrt{2}}\right) \exp \left( \frac{-q^{2}\ell ^{2}}{4}\right) ,
\label{K4}
\end{eqnarray}%
capture the character of the two different orbital states. Detailed
expressions for the Hartree and Fock interactions parameters are given in
Appendix A.

\section{ORDER PARAMETERS AT INTEGER FILLINGS}

The states with no pseudospin texture at integer filling factors have
uniform electronic density and density-matrices that vanish for $\mathbf{q}%
\neq 0$. Letting $\left\langle \rho _{n_{1},n_{2}}^{a,b}\left( \mathbf{q}%
=0\right) \right\rangle \rightarrow $ $\left\langle \rho
_{n_{1},n_{2}}^{a,b}\right\rangle ,$ the Hamiltonian of Eq. (\ref{hamil})
reduces to\qquad \qquad 
\begin{eqnarray}
H_{HF} &=&N_{\phi }E_{a,n}\rho _{n,n}^{a,a} \\
&&-N_{\phi }X_{n_{1},n_{4},n_{3},n_{2}}^{a,b}\left( 0\right) \left\langle
\rho _{n_{1},n_{2}}^{a,b}\right\rangle \rho _{n_{3},n_{4}}^{b,a}.  \notag
\end{eqnarray}%
These order parameters are conveniently calculated by defining the
time-ordered Matsubara Green's function

\begin{equation}
G_{n_{1},n_{2}}^{a,b}\left( \tau \right) =-\frac{1}{N_{\phi }}%
\sum_{X}\left\langle T_{\tau }c_{a,n_{1},X}\left( \tau \right)
c_{b,n_{2},X}^{\dagger }\left( 0\right) \right\rangle ,
\end{equation}%
since, at time zero, we have

\begin{equation}
G_{n_{1},n_{2}}^{a,b}\left( \tau =0^{-}\right) =\left\langle \rho
_{n_{2},n_{1}}^{b,a}\right\rangle .
\end{equation}

In the Hartree-Fock approximation, the equation of motion for the
single-particle Green's function is%
\begin{gather}
\left[ \hslash i\omega _{n}-\left( E_{a,n_{1}}-\mu \right) \right]
G_{n_{1},n_{2}}^{a,b}\left( \omega _{n}\right)  \label{iterationp} \\
+U_{n_{1},n_{3}}^{a,c}G_{n_{3},n_{2}}^{c,b}\left( \omega _{n}\right)
=\hslash \delta _{a,b}\delta _{n_{1},n_{2}},  \notag
\end{gather}%
where $\omega _{n}$ is a fermionic Matsubara frequency and 
\begin{equation}
U_{n_{1},n_{3}}^{a,c}=X_{n,n_{3},n_{1},n^{\prime }}^{a,c}\left( 0\right)
\left\langle \rho _{n,n^{\prime }}^{c,a}\right\rangle .
\end{equation}

The system of Eqs. (\ref{iterationp}) can be solved in an iterative way by
using some initial values for the parameters $\left\{ \left\langle \rho
_{n_{2},n_{1}}^{b,a}\right\rangle \right\} .$ In Ref. \onlinecite{BarlasPRL}%
, we solved this equation keeping valley, orbital, and spin indices. We
showed that the solutions of the Hartree-Fock equations for the balanced
bilayer ($\Delta _{B}=0$) follow a Hund's rules behavior. The spin
polarization is maximized first, then the layer polarization is maximized to
the greatest extent possible, and finally the orbital polarization is
maximized to the extent allowed by the first two rules. In the absence of
bias, the ordering of the first four states (with spin up) is given by 
\begin{eqnarray}
\left\vert S,0\right\rangle &=&\frac{1}{\sqrt{2}}\left\vert K,0\right\rangle
+\frac{1}{\sqrt{2}}\left\vert K^{\prime },0\right\rangle ,  \label{sas1} \\
\left\vert S,1\right\rangle &=&\frac{1}{\sqrt{2}}\left\vert K,1\right\rangle
+\frac{1}{\sqrt{2}}\left\vert K^{\prime },1\right\rangle , \\
\left\vert AS,0\right\rangle &=&\frac{1}{\sqrt{2}}\left\vert
K,0\right\rangle -\frac{1}{\sqrt{2}}\left\vert K^{\prime },0\right\rangle ,
\label{sas2} \\
\left\vert AS,1\right\rangle &=&\frac{1}{\sqrt{2}}\left\vert
K,1\right\rangle -\frac{1}{\sqrt{2}}\left\vert K^{\prime },1\right\rangle ,
\label{sas4}
\end{eqnarray}%
in this order. The next four states follow the same order but with spin
down. The occupation of these eight states are given by the filling factor $%
\nu $ ranging from $\nu =-3$ (state $\left\vert S,0\right\rangle $ with spin
up fully filled) to $\nu =+4$ (all eight states filled).

To simplify the notation, we define%
\begin{eqnarray}
\left\vert K,0\right\rangle &\rightarrow &\left\vert 1\right\rangle ,
\label{states} \\
\left\vert K,1\right\rangle &\rightarrow &\left\vert 2\right\rangle ,  \notag
\\
\left\vert K^{\prime },0\right\rangle &\rightarrow &\left\vert
3\right\rangle ,  \notag \\
\left\vert K^{\prime },1\right\rangle &\rightarrow &\left\vert
4\right\rangle ,  \notag
\end{eqnarray}%
so that $\left\langle \rho _{n_{2},n_{1}}^{b,a}\right\rangle \rightarrow
\left\langle \rho _{i,j}\right\rangle $ with $i,j=1,2,3,4.$

It is easy, using Eq. (\ref{iterationp}), to prove the sum rules%
\begin{equation}
\sum_{j=1}^{4}\left\vert \left\langle \rho _{i,j}\right\rangle \right\vert
^{2}=\nu _{i},
\end{equation}%
where 
\begin{equation}
\nu _{i}=\left\langle \rho _{i,i}\right\rangle .
\end{equation}

\section{COLLECTIVE MODES IN THE GENERALIZED RANDOM-PHASE APPROXIMATION}

In order to compute the collective excitations, we define the two-particle
Matsubara Green's function%
\begin{eqnarray}
&&\chi _{n_{1},n_{2},n_{3},n_{4}}^{a,b,c,d}\left( \mathbf{q},\tau \right) \\
&=&-N_{\phi }\left\langle T_{\tau }\rho _{n_{1},n_{2}}^{a,b}\left( \mathbf{q,%
}\tau \right) \rho _{n_{3},n_{4}}^{c,d}\left( -\mathbf{q},0\right)
\right\rangle  \notag \\
&&+N_{\phi }\left\langle \rho _{n_{1},n_{2}}^{a,b}\left( \mathbf{q}\right)
\right\rangle \left\langle \rho _{n_{3},n_{4}}^{c,d}\left( -\mathbf{q}%
\right) \right\rangle ,  \notag
\end{eqnarray}%
where again $n_{i}=0,1$ are orbital indices and $a,b,c,d$ are valley
indices. To derive the equation of motion for these response functions in
the Generalized Random-Phase Approximation (GRPA), we proceed in the
following way. We first derive the equation of motion for $\chi $ in the
Hartree-Fock Approximation (HFA) using the Heisenberg equation of motion 
\begin{equation}
\hslash \frac{\partial }{\partial \tau }\left( \ldots \right) =\left[ H-\mu
N,\left( \ldots \right) \right] ,
\end{equation}%
where $H$ is the Hamiltonian of Eq. (\ref{hamil}) with the averages removed, 
$\mu $ is the chemical potential and $N$ the number operator (not to be
confused with the Landau level index). After evaluating the commutators, we
linearize the resulting equation by writing $\rho \left( \mathbf{q}\right)
\rightarrow \left\langle \rho \left( \mathbf{q}\right) \right\rangle
_{HFA}+\delta \rho \left( \mathbf{q}\right) $. We get the GRPA\ equations of
motion by keeping the terms up to linear order in $\delta \rho \left( 
\mathbf{q}\right) .$ In the homogeneous states at integer fillings, $%
\left\langle \rho \left( \mathbf{q}\right) \right\rangle _{HF}=\left\langle
\rho \left( \mathbf{q}\right) \right\rangle _{HF}\delta _{\mathbf{q},0},$ so
that we get the set of equations 
\begin{eqnarray}
&&\left[ i\hslash \Omega _{n}-\left( E_{b,n_{2}}-E_{a,n_{1}}\right) \right]
\chi _{n_{1},n_{2},n_{3},n_{4}}^{\left( 0\right) a,b,c,d}\left( \mathbf{q}%
,\Omega _{n}\right)  \notag \\
&=&\hslash \left\langle \rho _{n_{1},n_{4}}^{a,d}\right\rangle \delta
_{n_{2},n_{3}}\delta _{b,c}-\hslash \left\langle \rho
_{n_{3},n_{2}}^{c,b}\right\rangle \delta _{a,d}\delta _{n_{1},n_{4}}
\label{chihfaa} \\
&&+X_{n_{1}^{\prime },n_{1},n_{3}^{\prime },n_{2}^{\prime }}^{a,b^{\prime
}}\left( 0\right) \left\langle \rho _{n_{1}^{\prime },n_{2}^{\prime
}}^{a,b^{\prime }}\right\rangle \chi _{n_{3}^{\prime
},n_{2},n_{3},n_{4}}^{\left( 0\right) b^{\prime },b,c,d}\left( \mathbf{q}%
,\Omega _{n}\right)  \notag \\
&&-X_{n_{1}^{\prime },n_{4}^{\prime },n_{2},n_{2}^{\prime }}^{a^{\prime
},b}\left( 0\right) \left\langle \rho _{n_{1}^{\prime },n_{2}^{\prime
}}^{a^{\prime },b}\right\rangle \chi _{n_{1},n_{4}^{\prime
},n_{3},n_{4}}^{\left( 0\right) a,a^{\prime },c,d}\left( \mathbf{q},\Omega
_{n}\right) ,  \notag
\end{eqnarray}%
and 
\begin{eqnarray}
&&\chi _{n_{1},n_{2},n_{3},n_{4}}^{a,b,c,d}\left( \mathbf{q};\Omega
_{n}\right)  \label{chigrpa} \\
&=&\chi _{n_{1},n_{2},n_{3},n_{4}}^{\left( 0\right) a,b,c,d}\left( \mathbf{q}%
;\Omega _{n}\right)  \notag \\
&&+\frac{1}{\hslash }\chi _{n_{1},n_{2},n_{5},n_{6}}^{\left( 0\right)
a,b,e,e}\left( \mathbf{q};\Omega _{n}\right)
H_{n_{5},n_{6},n_{7},n_{8}}^{e,g}\left( \mathbf{q}\right)  \notag \\
&&\times \chi _{n_{7},n_{8},n_{3},n_{4}}^{g,g,c,d}\left( \mathbf{q};\Omega
_{n}\right)  \notag \\
&&-\frac{1}{\hslash }\chi _{n_{1},n_{2},n_{5},n_{6}}^{\left( 0\right)
a,b,e,f}\left( \mathbf{q};\Omega _{n}\right)
X_{n_{5},n_{8},n_{7},n_{6}}^{e,f}\left( \mathbf{q}\right)  \notag \\
&&\times \chi _{n_{7},n_{8},n_{3},n_{4}}^{f,e,c,d}\left( \mathbf{q};\Omega
_{n}\right) ,  \notag
\end{eqnarray}%
where $\Omega _{n}$ is a bosonic Matsubara frequency. The retarded response
functions are obtained, as usual, by taking the analytic continuation $%
i\Omega _{n}\rightarrow \omega +i\delta .$

By defining super indices $A,B=1,2,3,...,16$ representing the combinations $%
(a,n_{1};b,n_{2}),\left( c,n_{3};d,n_{4}\right) ,$ etc., we can represent
the response functions and interactions matrices as $16\times 16$ matrices
and then write the GRPA\ equation in the matrix form:

\begin{equation}
\left[ \left( \omega +i\delta \right) I-F\left( \mathbf{q}\right) \right]
\chi \left( \mathbf{q},\omega \right) =B\left( \mathbf{q}\right) ,
\label{grpamodes}
\end{equation}%
where $B,I,F,\chi $ are $16\times 16$ matrices (with $I$ the unit matrix).
The matrices $F\left( \mathbf{q}\right) $ and $B\left( \mathbf{q}\right) $
depend on the $\left\langle \rho _{n_{2},n_{1}}^{b,a}\right\rangle ^{\prime
}s$ evaluated in the HFA. We will give later the precise form of these
matrices for the phases studied in this paper.

\begin{figure}[tbph]
\includegraphics[scale=1]{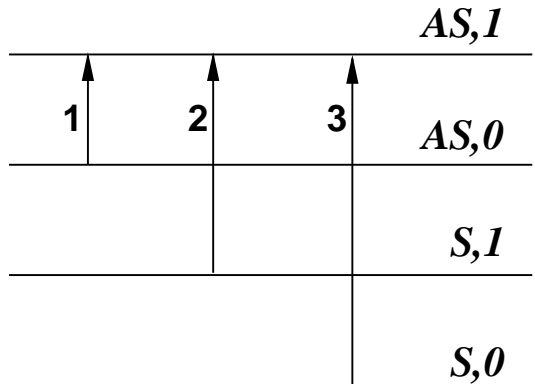}
\caption{ At $\protect\nu =-1$, the first three energy states of the
Hartree-Fock Hamiltonian are filled and the only possible transitions are
those indicated by the three arrows. The frequency of these transitions
corresponds to the frequency of the collective excitations calculated in the
GRPA\ in the limit in which the wavevector $q\rightarrow \infty .$ We
illustrate here the case of zero bias. }
\label{modes}
\end{figure}

The frequencies of the collective excitations are given by the eigenvalues
of the matrix $F\left( \mathbf{q}\right) .$ There are in total 4 zero modes,
corresponding to unphysical intra-level transitions, and 12 non-zero modes
corresponding to inter-level transitions. The latter occur in six
positive-negative energy pairs, corresponding to excitation and deexcitation
partners. Of the six collective excitations identified in this way, three
are Pauli-blocked at $\nu =-1$ and appear in our calculations as
dispersionless modes that have zero weight in all physical response
properties. The three remaining excitation modes are physical and for $\nu
=-1$ correspond to the interaction-coupled transitions indicated in Fig. \ref%
{modes}. Note that, in the limit $q\rightarrow \infty $, $%
H^{e,g},X^{e,f}\rightarrow 0$ so that Eq. (\ref{chigrpa}) gives $\chi
\rightarrow \chi ^{0}$. In this limit, the collective mode frequencies
correspond to transitions between eigenstates of the Hartree-Fock
Hamiltonian $H_{HF}$ as illustrated in Fig. \ref{modes} for the case $\nu
=-1 $ and zero bias. The energy of these eigenstates include the
non-interacting energies and the self-energy corrections. The $\omega \left( 
\mathbf{q}=0\right) $ limit of these sames modes, however, also includes the
polarization and excitonic corrections that, in a Feynman diagram
description of the GRPA, are captured by bubble and ladder diagram
summations. These effects make the modes dispersive.

\section{PHASE DIAGRAM AT FILLING FACTOR $\protect\nu =-1$}

The properties of the ground state at $\nu =-3$ have been studied in detail
in Ref. \onlinecite{yafis2}. At zero bias, the ground state has all
electrons in the $\left\vert S,0\right\rangle $ state and can be described
as an $XY$ layer-pseudospin ferromagnet with orbital character $n=0$. The
bias $\Delta _{B}$ acts as an effective external magnetic field that forces
the layer pseudospins out of the $x-y$ plane. Above a critical bias $\Delta
_{B}^{\left( c\right) }$, all electrons are in the bottom layer and the
layer-pseudospin is correspondingly fully polarized. The ground state is
then given by $\left\vert K^{\prime },0\right\rangle $ and is unchanged if
the bias is further increased. The $\nu =-1$ state on which we focus here
differs from the $\nu =-3$ state because of a competition between
single-particle and interaction energy effects which emerges only in the
former case. As illustrated in Fig. (\ref{niveaux}), single-particle effects
captured by the bilayer effective Hamiltonian favor occupation of the $n=1$
orbital when three of the octet's eight levels are occupied ($\nu =-1$).
Note that this tendency is independent of the sign of $\Delta _{B}$.
Exchange interactions, on the other hand, always favor a state in which as
many $n=0$ orbitals as possible are occupied. As we explain below, a
compromise is reached by forming a state with coherence between $n=0$ and $%
n=1$ orbitals. This physics is enriched by the same tendency toward
interlayer coherence which occurs at $\nu =-3$ and\cite{coherencerefs} in
semiconductor bilayers. Indeed, our calculations show that the phase diagram
at $\nu =-1$ is much more complex than at $\nu =-3$.

\subsection{Inter-layer-coherent state}

\textit{At zero bias}, numerical solution of the HFA equations leads to
occupied $\left\vert S,0\right\rangle ,\left\vert S,1\right\rangle ,$ and $%
\left\vert AS,0\right\rangle $ states. The order parameters are then given
by 
\begin{eqnarray}
\nu _{1} &=&\nu _{3}=1,  \label{phase} \\
\nu _{2} &=&\nu _{4}=\frac{1}{2},  \notag \\
\left\langle \rho _{2,4}\right\rangle &=&\left\langle \rho
_{4,2}\right\rangle =\frac{1}{2}.  \notag
\end{eqnarray}%
The ground state is an $XY$ layer-pseudospin ferromagnet with orbital
character $n=1.$ The Hamiltonian is invariant with respect to the
orientation of the pseudospins in the $x-y$ plane so that this phase
supports a Goldstone mode. The choice of phase in Eq. (\ref{phase}) has the
pseudospins pointing along the $x$ axis. At finite bias $\Delta _{B}<\Delta
_{B}^{(1)}$, the pseudospins are pushed out the $x-y$ plane i.e. the two
layers have unequal population. The occupation of the four states are in
this case%
\begin{equation}
\nu _{1}=\nu _{3}=1,
\end{equation}%
and%
\begin{equation}
\nu _{2}=1-\nu _{4}=\frac{1}{2}\left( 1-\frac{\Delta _{B}}{\Delta
_{B}^{\left( 1\right) }}\right) ,
\end{equation}%
with inter-layer coherence reflected by 
\begin{equation}
\left\langle \rho _{2,4}\right\rangle =\left\langle \rho _{4,2}\right\rangle
=\sqrt{\nu _{2}-\nu _{2}^{2}}.
\end{equation}

We define the critical bias $\Delta _{B}^{\left( 1\right) }$ as the bias at
which the inter-layer coherence $\left\langle \rho _{2,4}\right\rangle =0.$
It is given by%
\begin{equation}
\Delta _{B}^{\left( 1\right) }=\frac{-3x_{7}\left( 0\right) +2\widetilde{x}%
_{16}\left( 0\right) +2\frac{d}{\ell }}{2-4\beta },
\end{equation}%
where the Fock interactions $x_{i}\left( \mathbf{q}\right) $ and $\widetilde{%
x}_{i}\left( \mathbf{q}\right) $ are defined in Appendix A. As an example,
for $B=10$ T, $\Delta _{B}^{\left( 1\right) }=0.00205\,e^{2}/\kappa \ell .$

The energy of this inter-layer-coherent state (ICS)\ for $\Delta _{B}<\Delta
_{B}^{(1)}$ is 
\begin{eqnarray}
\frac{E_{HF}^{(1)}}{N} &=&\frac{1}{3}\Delta _{B}\left( 1-2\beta \right)
\left\langle P_{z,1}\right\rangle  \label{ener1} \\
&&-\frac{1}{3}x_{1}\left( 0\right) -\frac{1}{3}x_{7}\left( 0\right) -\frac{1%
}{12}x_{16}\left( 0\right)  \notag \\
&&+\frac{1}{3}\left( \frac{d}{\ell }-x_{16}\left( 0\right) \right)
\left\langle P_{z,1}\right\rangle ^{2}  \notag \\
&&-\frac{1}{3}\widetilde{x}_{16}\left( 0\right) \left\langle \mathbf{P}%
_{\bot ,1}\right\rangle ^{2},  \notag
\end{eqnarray}%
where the components of the layer pseudospin are given by%
\begin{eqnarray}
\left\langle P_{z,1}\right\rangle &=&\frac{\nu _{2}-\nu _{4}}{2}=-\frac{1}{2}%
\frac{\Delta _{B}}{\Delta _{B}^{\left( 1\right) }}, \\
\left\langle \mathbf{P}_{\bot ,1}\right\rangle ^{2} &=&\left\langle
P_{x,1}\right\rangle ^{2}+\left\langle P_{y,1}\right\rangle ^{2}=\left\vert
\left\langle \rho _{2,4}\right\rangle \right\vert ^{2},
\end{eqnarray}%
with the convention that pseudospin up is state $\left\vert 2\right\rangle $
while pseudospin down is state $\left\vert 4\right\rangle .$

\subsection{Inter-orbital-coherent state}

The HFA equations have a separate set of solutions, favored at larger bias
voltages, in which the lower layer is maximally occupied, and the ground
state has upper-layer orbital coherence instead of inter-layer coherence.
This solution has 
\begin{equation}
\nu _{3}=\nu _{4}=1,  \label{m1}
\end{equation}%
\begin{equation}
\nu _{2}=1-\nu _{1}=\frac{\Delta _{B}}{\Delta _{B}^{(2)}},  \label{m2}
\end{equation}%
and inter-orbital coherence is signalled by the density-matrix components 
\begin{equation}
\left\langle \rho _{1,2}\right\rangle =\left\langle \rho _{2,1}\right\rangle
=\sqrt{\nu _{2}-\nu _{2}^{2}}.  \label{m3}
\end{equation}%
Here 
\begin{equation}
\Delta _{B}^{(2)}=\frac{x_{4}\left( 0\right) }{2\beta }
\end{equation}%
is the critical bias above which all charges in the upper layer are
transferred to state $\left\vert 2\right\rangle $ and the orbital coherence
is lost. In a pseudospin model with the convention: pseudospin up for state $%
\left\vert 1\right\rangle $ and pseudospin down for state $\left\vert
2\right\rangle $, the orbital pseudospin components are given by 
\begin{eqnarray}
\left\langle S_{z,K}\right\rangle &=&\frac{\nu _{1}-\nu _{2}}{2}, \\
\left\langle \mathbf{S}_{\bot ,1}\right\rangle &=&\left\langle
S_{x,1}\right\rangle \widehat{\mathbf{x}}+\left\langle S_{y,1}\right\rangle 
\widehat{\mathbf{y}}, \\
\left\langle S_{+,K}\right\rangle &=&\left\langle S_{x,K}\right\rangle
+i\left\langle S_{y,K}\right\rangle =\left\langle \rho _{1,2}\right\rangle .
\end{eqnarray}%
This orbital-coherent phase has all pseudospins tilted slightly away from
the $z$ axis by an angle%
\begin{equation}
\cos \theta _{B}=1-2\frac{\Delta _{B}}{\Delta _{B}^{(2)}}.  \label{angle}
\end{equation}%
At the critical bias $\Delta _{B}^{(2)}$, $\theta _{B}=\pi .$ This critical
bias $\Delta _{B}^{(2)}$ is very large; at $B=10$ T, $\Delta
_{B}^{(2)}\approx 5e^{2}/\kappa \ell $ which is near the limit of validity
of the effective two-band model i.e. $\hslash \omega _{c}<\gamma _{1}$. For $%
\Delta _{B}>\Delta _{B}^{(2)}$, all electrons are in state $\left\vert
2\right\rangle $ and there is no further change with bias of the ground
state.

The orbital-coherent state (OCS) has an energy given by%
\begin{eqnarray}
\frac{E_{HF}^{(2)}}{N} &=&\frac{1}{3}\left[ -\left( 1-\beta \right) \frac{%
\Delta _{B}}{2}+\beta \Delta _{B}\left\langle S_{z,K}\right\rangle \right] +%
\frac{1}{12}\frac{d}{\ell }  \notag \\
&&-\frac{5}{24}\left[ x_{1}\left( 0\right) +x_{16}\left( 0\right)
+2x_{7}\left( 0\right) \right]  \label{ener22} \\
&&-\frac{1}{6}\left[ x_{1}\left( 0\right) -x_{16}\left( 0\right) \right]
\left\langle S_{z,K}\right\rangle  \notag \\
&&-\frac{1}{6}\left[ x_{1}\left( 0\right) +x_{16}\left( 0\right)
-4x_{7}\left( 0\right) \right] \left\langle S_{z,K}\right\rangle ^{2}  \notag
\\
&&-\frac{1}{3}x_{4}\left( 0\right) \left\langle \mathbf{S}_{\bot
,K}\right\rangle ^{2}.  \notag
\end{eqnarray}%
This energy is independent of the azimuthal angle $\varphi $ of the
pseudospin vector. We can thus, without loss of generality, take $%
\left\langle \rho _{1,2}\right\rangle $ as real.

\begin{figure}[tbph]
\includegraphics[scale=1]{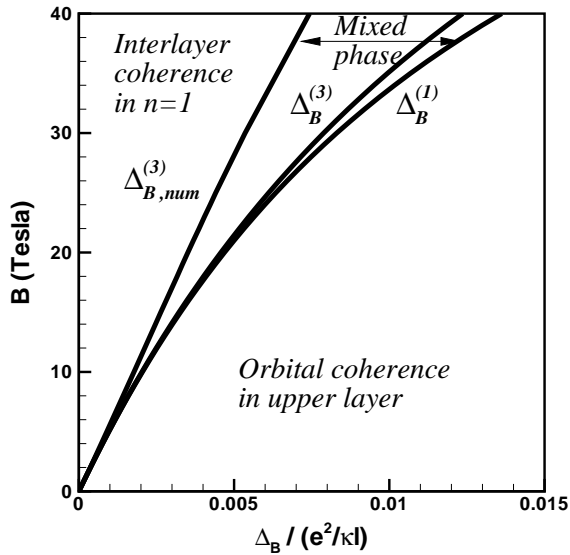}
\caption{ Phase diagram of the 2DEG\ in a graphene bilayer at $\protect\nu %
=-1.$ }
\label{biasnum1}
\end{figure}

\subsection{Mixed state}

Eqs. (\ref{ener1}) and (\ref{ener22}) give $E_{HF}^{(2)}<E_{HF}^{(1)}$ above
a critical bias $\Delta _{B}^{(3)}<\Delta _{B}^{(1)}$. For biases $\Delta
\in \left[ \Delta _{B}^{(3)},\Delta _{B}^{(1)}\right] ,$ a mixed state with
both inter-orbital \emph{and} inter-layer coherence would be lower in energy
than a state with only interlayer coherence. Solving the full Hartree-Fock
equations, we find that the crossover from the inter-layer coherent state to
the inter-orbital coherent state occurs continuously via an intermediate
state with both orders. The boundaries of the intermediate phase must be
determined numerically. We find that the boundaries of this mixed state (MS)
are given on the left by a new critical bias $\Delta _{B,num}^{(3)}<\Delta
_{B}^{(3)}$ and on the right by $\Delta _{B}^{(1)}.$ The intermediate phase
region broadens with magnetic field as can be seen in Fig. \ref{biasnum1}.

\begin{figure}[tbph]
\includegraphics[scale=1]{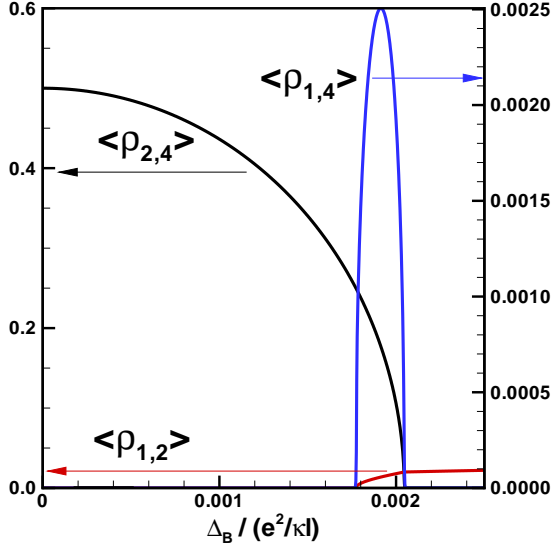}
\caption{ (Color online) Evolution of the inter-layer $\left\langle \protect%
\rho _{2,4}\right\rangle $ and inter-orbital $\left\langle \protect\rho %
_{1,2}\right\rangle $ coherences (left axis) with bias at filling factor $%
\protect\nu =-1$ and magnetic field $B=10$ T. On the right axis, one of the
non zero coherences $\left\langle \protect\rho _{1,4}\right\rangle $ in the
mixed state. }
\label{pordre}
\end{figure}

Figure \ref{pordre} shows the evolution of the inter-layer coherence $%
\left\langle \rho _{2,4}\right\rangle $ and the orbital coherence $%
\left\langle \rho _{1,2}\right\rangle $ with bias at magnetic field $B=10$
T. The orbital coherence sets in before the inter-layer coherence decreases
to zero thus creating the mixed state region identified in this figure by a
non zero value of the density-matrix component $\left\langle \rho
_{1,4}\right\rangle $. (Note that $\left\langle \rho _{1,2}\right\rangle $
is not given by Eq. (\ref{m3}) in the mixed state.)\ The coherences $%
\left\langle \rho _{1,4}\right\rangle $ and $\left\langle \rho
_{2,3}\right\rangle ,$ which involve a mixing of valley and as well as
orbital indices, are non-zero only in the intermediate mixed-state region of
the phase diagram. All density-matrix components vary continuously with
inter-layer bias.

As can be seen from Figs. \ref{biasnum1} and \ref{pordre}, the
orbital-coherent phase starts at $\Delta _{B}^{(1)}$ with a finite orbital
coherence $\left\langle \rho _{1,2}\right\rangle $. Were it not for the
presence of the inter-layer-coherent and mixed states, $\left\langle \rho
_{1,2}\right\rangle $ would start at zero bias and be given by Eq. (\ref{m3}%
) for all biaises. The mixed and inter-layer coherent states are confined to
relatively small inter-layer bias voltages; at larger values of $\Delta _{B}$
the ground state is a relatively simple state with only orbital coherence.
The exploration of collective excitation properties of this state is one key
objective of this paper.

\section{COLLECTIVE MODES IN THE INTER-LAYER COHERENT STATE}

The density-matrix equations of motions which describe the three $\nu =-1$
spin-diagonal dispersive collective modes (see Fig. \ref{modes}) are usually
simplest when written in the basis of the bonding and antibonding
single-particle Hartree-Fock eigenstates. For the interlayer coherent state
(ICS) (the ground state for $\Delta _{B}<$ $\Delta _{B,num}^{(3)}$) we find
that, 
\begin{equation}
\left( I\left( \omega +i\delta \right) -F_{1}\right) \left( 
\begin{array}{c}
ie^{-i\theta _{\mathbf{q}}}\rho _{AB1,B0} \\ 
ie^{i\theta _{\mathbf{q}}}\rho _{B0,AB1} \\ 
\rho _{AB1,B1} \\ 
\rho _{B1,AB1} \\ 
ie^{i\theta _{\mathbf{q}}}\rho _{AB0,AB1} \\ 
ie^{-i\theta _{\mathbf{q}}}\rho _{AB1,AB0}%
\end{array}%
\right) =\left( 
\begin{array}{c}
0 \\ 
0 \\ 
0 \\ 
0 \\ 
0 \\ 
0%
\end{array}%
\right) ,  \label{f1q}
\end{equation}%
where $\theta _{\mathbf{q}}$ is the angle between the two-dimensional
wavevector $\mathbf{q}$ and the $x$ axis, and $B$ and $AB$ refer to the
states 
\begin{eqnarray}
\left\vert B,n,X\right\rangle &=&g_{-}\left\vert K,n,X\right\rangle
+g_{+}\left\vert K^{\prime },n,X\right\rangle , \\
\left\vert AB,n,X\right\rangle &=&g_{+}\left\vert K,n,X\right\rangle
-g_{-}\left\vert K^{\prime },n,X\right\rangle ,
\end{eqnarray}%
where $n=0,1,$%
\begin{equation}
g_{\pm }=\sqrt{\frac{1\pm \sigma }{2}},
\end{equation}%
and%
\begin{equation}
\sigma =\frac{\Delta _{B}}{\Delta _{B}^{\left( 1\right) }}.
\end{equation}%
The matrix $F_{1}$ depends only on the modulus of the wavevector $\mathbf{q}$
so that the dispersions are isotropic. This matrix is given by

\begin{widetext}
\begin{equation}
F_{1}\left( q\right) =\frac{1}{2}\left( 
\begin{array}{cccccc}
-A-\sigma ^{2}B & fH & -N+\sigma ^{2}J & -fJ & \sigma\sqrt{f}H & -\sigma\sqrt{f}G \\ 
-fH & A+\sigma ^{2}B & fJ & N-\sigma ^{2}J & \sigma\sqrt{f}G & -\sigma\sqrt{f}H \\ 
-N+\sigma ^{2}J & -fJ & -C-\sigma ^{2}D & fM & -\sigma\sqrt{f}J & -\sigma\sqrt{f}J \\ 
fJ & N-\sigma ^{2}J & -fM & C+\sigma ^{2}D & \sigma\sqrt{f}J & \sigma\sqrt{f}J \\ 
-\sigma\sqrt{f}H & \sigma\sqrt{f}G & \sigma\sqrt{f}J & \sigma\sqrt{f}J & -E-\sigma ^{2}F & K-\sigma
^{2}H \\ 
-\sigma\sqrt{f}G & \sigma\sqrt{f}H & -\sigma\sqrt{f}J & -\sigma\sqrt{f}J & -K+\sigma ^{2}H & 
E+\sigma ^{2}F%
\end{array}%
\right) ,  \label{f1}
\end{equation}%
\end{widetext}

where%
\begin{equation}
f=1-\sigma ^{2}
\end{equation}%
The variables in this matrix are defined in Appendix B.

\begin{figure}[tbph]
\includegraphics[scale=1]{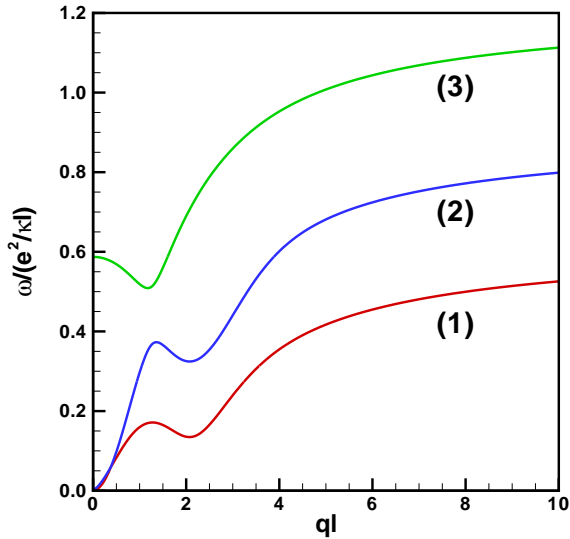}
\caption{ (Color online) Dispersion relations of the collective modes at $%
\protect\nu =-1$ in the inter-layer coherent phase at zero bias and for $%
B=10 $ T. The numbers refer to the transitions indicated in Fig. \protect\ref%
{modes}. }
\label{modesnum1}
\end{figure}

The GRPA dispersions for the three collective modes are shown in Fig. \ref%
{modesnum1} for zero bias and a magnetic field of $B=10$ T. In the limit $%
q\rightarrow \infty $, the frequencies of the dispersive modes correspond to
transitions between the HFA energy levels indicated in Fig. \ref{modes} as
expected. Mode $2$ in Fig. \ref{modesnum1} is a Goldstone mode consisting of
a precession of the inter-layer pseudospin $\mathbf{P}_{1}$ around the $x$
axis. We refer to it as the inter-layer pseudospin mode (IPM). Mode $1$ is
an orbital pseudospin mode (OPM)\ consisting of a precession of the orbital
pseudospins around their local\ equilibrium position. Mode $3$ involves both
a layer and orbital pseudospin flip and has a large gap. At finite bias, the 
$3$ dispersive modes in Fig. \ref{modes} are coupled together while at zero
bias, modes $2$ and $3$ completely decouple from mode $1$ as is clear from
Eq. (\ref{f1}).

\begin{figure}[tbph]
\includegraphics[scale=1]{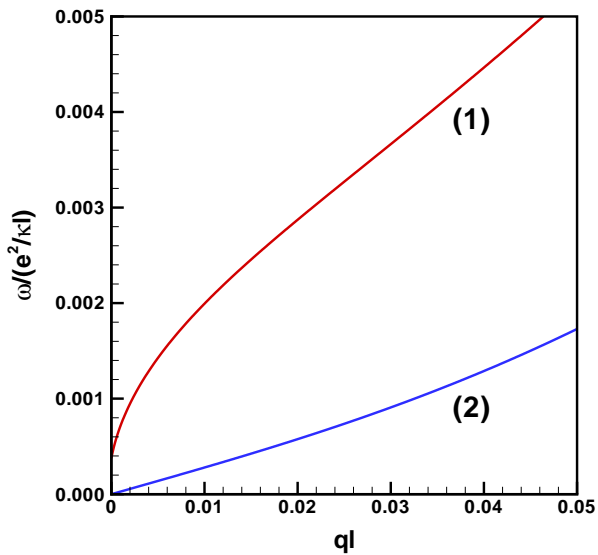}
\caption{ (Color online) Long-wavelength (small $q$) dispersion relation of
the collective modes in the inter-layer coherent phase at $\protect\nu =-1$
for zero bias and for $B=10$ T. Mode $1$ is the orbital pseudospin mode
while mode $2$ is the inter-layer pseudospin mode. }
\label{modesnum1petitq}
\end{figure}

From Eq. (\ref{f1}), it is easy to see that the dispersion of the OPM\ at
zero bias (given by the $2\times 2$ block in the lower right of the matrix $%
F_{1}$) is given by%
\begin{equation}
\omega _{OPM}\left( \mathbf{q}\right) =\frac{1}{2}\sqrt{E^{2}\left( \mathbf{q%
}\right) -K^{2}\left( \mathbf{q}\right) },
\end{equation}%
and has a gap given by (see Appendix B)%
\begin{equation}
\omega _{OPM}\left( 0\right) =\frac{1}{2}\left[ x_{1}\left( 0\right) -%
\widetilde{x}_{1}\left( 0\right) -x_{16}\left( 0\right) +\widetilde{x}%
_{16}\left( 0\right) \right] .  \label{gaporbi}
\end{equation}%
This gap is small but visible in Fig. \ref{modesnum1petitq}. We find
numerically, that as the bias is increased, the gap $\omega _{OPM}\left(
0\right) $ decreases until it reaches zero at the phase boundary of the
mixed state.

The dispersion of the IPM is linear in $q$ for $q\ell <d/\ell $ (with $%
d/\ell \approx 0.04$ for $B=10$T) as in a semiconductor bilayer\cite%
{spielmanPRL}. This findings differ qualitatively from the $\nu =-3,$ case
for which we found\cite{yafis2} an IPM with $q^{2}$ dispersion and a gapless
OPM. At $\nu =-3$ and zero bias, the $\left\vert S,0\right\rangle $ level is
filled. The $q^{2}$ IPM dispersion in that case occurs because the
possibility of mixing $n=1$ wavefunctions with $\left\vert AS,0\right\rangle 
$ wavefunctions in excited states allows the inter-layer phase stiffness to
vanish. At $\nu =-1$, it is not possible to make the corresponding admixture.

In pseudospin language, a finite bias pushes the layer pseudospins $\mathbf{P%
}_{1}$ out of the $x-y$ plane but the Hamiltonian of the system remains
independent of the orientation of the perpendicular component of these
pseudospins in the $x-y$ plane. The IPM therefore remains gapless for $%
\Delta _{B}<\Delta _{B,num}^{(3)}.$ It acquires a gap in the mixed and OCS.
At filling factor $\nu =-3$, the inter-layer-pseudospin mode becomes
unstable at finite bias. This indicates that the uniform
inter-layer-coherent state cannot in fact be the ground state at finite
bias. We see no such instability at $\nu =-1.$

In the absence of a bias, Eq. (\ref{f1}) shows that modes $2$ and $3$ are
coupled through $N$ and $J.$ These interactions involves the Coulomb
interaction matrix elements $\widehat{X}_{0,1,1,1}\left( \mathbf{q}\right) $
and $\widehat{H}_{0,1,1,1}\left( \mathbf{q}\right) $ (see Appendix A for
their definitions). These interactions do not conserve total $n=0$ or $n=1$
quantum numbers. Such interactions do not occur in usual semiconductor 2DEG
where spin and layer pseudospin indices are conserved.

\section{COLLECTIVE MODES IN THE ORBITAL COHERENT STATE}

In this section, we consider collective excitations of the orbital coherent
state (OCS) which is the ground state in the region $\Delta _{B}^{\left(
1\right) }<\Delta _{B}<\Delta _{B}^{(2)}$, which covers the large region of
bias voltages from small values to the largest values for which the two-band
effective model applies. The occurrence of the interesting OCS state at high
bias voltages is a consequence of competition between single-particle and
interaction effects as explained earlier. By studying its collective
interactions we reveal a Dzyaloshinskii-Moriya (DM) interaction between
orbital pseudospins and demonstrate that for large bias voltages it drives
an instability to an orbital pseudospin spiral state.

\subsection{Electric dipole density}

The fact that $\left\langle \rho _{1,2}\right\rangle \neq 0$ in the OCS
implies that there is a finite density of electric dipoles in this phase as
first pointed out in Ref. \onlinecite{shizuya1}. To show this, we write the
total electronic density (including the two valleys) as%
\begin{equation}
n\left( \mathbf{q}\right) =n_{K}\left( \mathbf{q}\right) +n_{K^{\prime
}}\left( \mathbf{q}\right) ,
\end{equation}%
where%
\begin{equation}
n_{K}\left( \mathbf{q}\right) =N_{\varphi }\sum_{i,j=0,1}K_{i,j}\left( -%
\mathbf{q}\right) \rho _{i,j}^{K,K}\left( \mathbf{q}\right) ,
\end{equation}%
with $\rho _{0,1}^{K,K}=\rho _{1,2}$. The functions $K_{i,j}\left( \mathbf{q}%
\right) $ are defined in Eqs. (\ref{K1}-\ref{K4}). In our pseudospin langage
for the orbital states, we have the relations (for the $K$ valley) 
\begin{eqnarray}
\rho ^{K}\left( \mathbf{q}\right) &=&\rho _{0,0}^{K,K}\left( \mathbf{q}%
\right) +\rho _{1,1}^{K,K}\left( \mathbf{q}\right) , \\
\rho _{z}^{K}\left( \mathbf{q}\right) &=&\frac{1}{2}\left[ \rho
_{0,0}^{K,K}\left( \mathbf{q}\right) -\rho _{1,1}^{K,K}\left( \mathbf{q}%
\right) \right] , \\
\rho _{+}^{K}\left( \mathbf{q}\right) &=&\rho _{0,1}^{K,K}\left( \mathbf{q}%
\right) =\rho _{x}^{K}\left( \mathbf{q}\right) +i\rho _{y}^{K}\left( \mathbf{%
q}\right)
\end{eqnarray}%
and so the density operator in the $K$ valley can be written as%
\begin{eqnarray}
n_{K}\left( \mathbf{q}\right) &=&N_{\varphi }\left( 1-\frac{q^{2}\ell ^{2}}{4%
}\right) \overline{\rho }_{K}\left( \mathbf{q}\right) \\
&&+N_{\varphi }\left( \frac{q^{2}\ell ^{2}}{2}\right) \overline{\rho }%
_{K,z}\left( \mathbf{q}\right)  \notag \\
&&-N_{\varphi }\sqrt{2}iq_{x}\ell \;\overline{\rho }_{K,x}\left( \mathbf{q}%
\right)  \notag \\
&&+N_{\varphi }\sqrt{2}iq_{y}\ell \;\overline{\rho }_{K,y}\left( \mathbf{q}%
\right) ,  \notag
\end{eqnarray}%
where we have defined $\overline{\rho }_{K}\left( \mathbf{q}\right) =\exp
\left( -q^{2}\ell ^{2}/4\right) \rho ^{K}\left( \mathbf{q}\right) $ and
similarly for $\overline{\rho }_{K,i}$ with $i=x,y,z.$

Now, if the 2DEG is in an external electric field $\mathbf{E}\left( \mathbf{r%
}\right) =-\nabla \phi \left( \mathbf{r}\right) ,$ we have for the coupling
Hamiltonian%
\begin{eqnarray}
H_{ext} &=&-e\int d\mathbf{r}\;n\left( \mathbf{r}\right) \phi \left( \mathbf{%
r}\right) \\
&=&-\frac{e}{S}N_{\varphi }\sum_{j=K,K^{\prime }}\sum_{\mathbf{q}}e^{i%
\mathbf{q}\cdot \mathbf{r}}\left[ \left( 1-\frac{q^{2}\ell ^{2}}{4}\right) 
\overline{\rho }_{j}\left( -\mathbf{q}\right) \right.  \notag \\
&&+\left( \frac{q^{2}\ell ^{2}}{2}\right) \overline{\rho }_{j,z}\left( -%
\mathbf{q}\right)  \notag \\
&&\left. -\sqrt{2}i\left( q_{x}\ell \;\overline{\rho }_{j,x}\left( -\mathbf{q%
}\right) +q_{y}\ell \;\overline{\rho }_{j,y}\left( -\mathbf{q}\right)
\right) \right] \phi \left( \mathbf{q}\right) .  \notag
\end{eqnarray}%
With the electric field in the plane of the 2DEG, this coupling can be
written, in real space, as%
\begin{eqnarray}
H_{ext} &=&-eN_{\varphi }\sum_{j=K,K^{\prime }}\int d\mathbf{r}\left[ 
\overline{\rho }_{j}\left( \mathbf{r}\right) \phi \left( \mathbf{r}\right)
\right.  \label{dipole} \\
&&-\left. \frac{1}{4}\left( \overline{\rho }_{j}\left( \mathbf{r}\right)
\ell ^{2}-2\overline{\rho }_{j,z}\left( \mathbf{r}\right) \ell ^{2}\right)
\left( \mathbf{\nabla }\cdot \mathbf{E}\left( \mathbf{r}\right) \right) %
\right]  \notag \\
&&+\sqrt{2}\ell \int d\mathbf{r}\left[ \overline{\rho }_{j,x}\left( \mathbf{r%
}\right) E_{x}\left( \mathbf{r}\right) -\overline{\rho }_{j,y}\left( \mathbf{%
r}\right) E_{y}\left( \mathbf{r}\right) \right] .  \notag
\end{eqnarray}%
The inter-orbital coherence is zero in the $K^{\prime }$ valley and so $%
\left\langle \overline{\rho }_{K^{\prime },x}\right\rangle ,\left\langle 
\overline{\rho }_{K^{\prime },y}\right\rangle =0.$ Eq. (\ref{dipole})
implies that the dipole density in the 2D plane is%
\begin{equation}
\left\langle \overline{\mathbf{d}}_{K}\left( \mathbf{r}\right) \right\rangle
=-e\sqrt{2}\ell N_{\varphi }\left( \left\langle \overline{\rho }_{K,x}\left( 
\mathbf{r}\right) \right\rangle \widehat{\mathbf{x}}-\left\langle \overline{%
\rho }_{K,y}\left( \mathbf{r}\right) \right\rangle \widehat{\mathbf{y}}%
\right) ,  \label{dipoled}
\end{equation}%
or%
\begin{eqnarray}
\left\langle \mathbf{d}_{K}\left( \mathbf{q}\right) \right\rangle &=&-e\sqrt{%
2}\ell N_{\varphi }e^{-q^{2}\ell ^{2}/4}  \label{dipoleq} \\
&&\times \left( \left\langle \rho _{K,x}\left( \mathbf{q}\right)
\right\rangle \widehat{\mathbf{x}}-\left\langle \rho _{K,y}\left( \mathbf{q}%
\right) \right\rangle \widehat{\mathbf{y}}\right) .  \notag
\end{eqnarray}

The orientation of the dipole density is set by the phase of the
density-matrix component $\left\langle \rho _{0,1}^{K,K}\left( \mathbf{r}%
\right) \right\rangle =$ $\left\langle \rho _{1,2}\left( \mathbf{r}\right)
\right\rangle $ which specifies the phase of the spontaneously established
coherence between $n=0$ and $n=1$ orbitals in the ground state. For our
choice of the spontaneously established phase of $\left\langle \rho
_{1,2}\left( \mathbf{r}\right) \right\rangle $ in Eq. (\ref{m3}), the
dipoles are oriented along the $x$ axis. Because the $K^{\prime }$ valley
(bottom layer) Landau levels are maximally filled, there is no inter-orbital
coherence and therefore no contribution to the electric-dipole density from
the $\mathbf{K}^{\prime }$ valley.

\subsection{Effective pseudospin model}

\begin{figure}[tbph]
\includegraphics[scale=1]{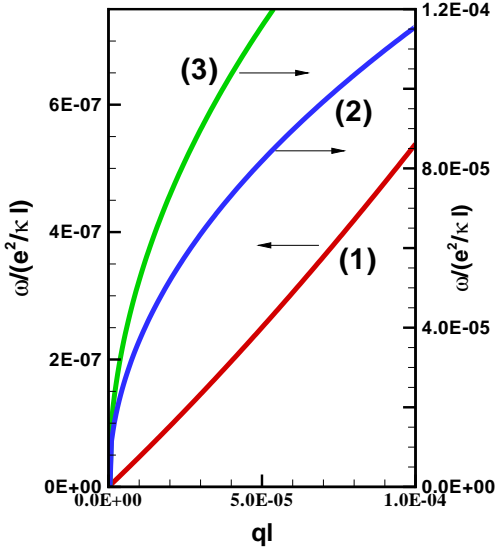}
\caption{ (Color online) Dispersion relations of the collective modes in the
orbital-coherent phase at filling factor $\protect\nu =-1$, bias $\Delta
_{B}=0.0022$ $e^{2}/\protect\kappa \ell $ and magnetic field $B=10$ T. Curve
(1) $q_{y}=0;$ curve (2) $q_{x}=q_{y};$ and curve (3) $q_{x}=0.$ }
\label{petitsq}
\end{figure}

Collective modes dispersions for the OCS are plotted in Fig. \ref%
{3modesgrandq}. For relatively small values of $\Delta _{B}$ they are very
similar to those represented in Fig. \ref{modesnum1} for the ICS; the main
changes occur at small wavevector as can be seen in the inset of Fig. \ref%
{3modesgrandq}. The inter-layer pseudospin mode is gapped in the OCS while
the orbital pseudospin mode is gapless, with a very anisotropic dispersion
as shown in Fig. \ref{petitsq}. We now discuss the physics of the orbital
pseudospin mode.

\begin{figure}[tbph]
\includegraphics[scale=1]{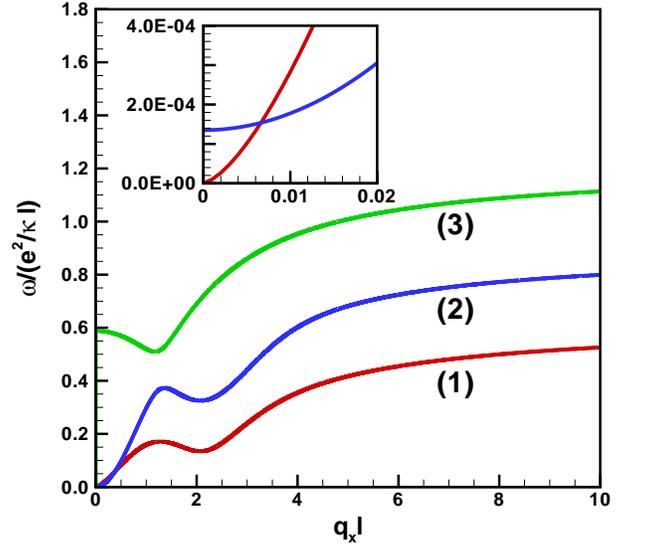}
\caption{ (Color online) Dispersion relations of the collective modes along $%
q_{x}$ in the orbital-coherent state at filling factor $\protect\nu =-1,$
bias $\Delta _{B}=0.0022$ $e^{2}/\protect\kappa \ell $ and magnetic field $%
B=10$ T. Mode $1$ is the Goldstone mode due to orbital coherence. Modes $2$
and $3$ involve an inter-layer transition and are gapped. The inset shows
the dispersion of mode $1$ (gapless) and mode $2$ (gapped) at small $%
q_{x}\ell .$ }
\label{3modesgrandq}
\end{figure}

At finite wavevector $\mathbf{q}$, the orbital pseudospin mode (OPM)
corresponds to a precession of the orbital pseudospins around their
equilibrium orientation in the ground state as illustrated in Fig. \ref%
{blochs}. The ground state is described by spinors 
\begin{equation}
\left\vert K,B,X\right\rangle =\cos \left( \frac{\theta _{B}}{2}\right)
\left\vert K,0,X\right\rangle +\sin \left( \frac{\theta _{B}}{2}\right)
\left\vert K,1,X\right\rangle ,
\end{equation}%
where $\theta _{B}\in \left[ 0,\pi \right] $ has been defined in Eq. (\ref%
{angle}).

\begin{figure}[tbph]
\includegraphics[scale=0.35]{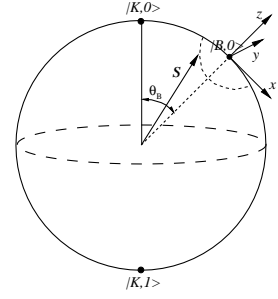}
\caption{Precession of the orbital pseudospin $\mathbf{S}$ around its
equilibrium position with polar angle $\protect\theta _{B}$ and azimuthal
angle $\protect\varphi _{B}=0$. This orientation is choosen as the orbital
pseudospin quantization axis in the pseudospin wave model. The angle $%
\protect\theta $ is the polar angle measured from the equilibrium
orientation of the pseudospins.}
\label{blochs}
\end{figure}

The collective mode corresponds to spatially coherent rotations of this
spinor around its ground state value. For this reason, it is convenient to
choose the orbital pseudospin quantization axis along the direction $%
\left\vert B,0\right\rangle $. In order to do this, we define `bonding' and
`antibonding' electron creation operators by

\begin{eqnarray}
b_{B,K,X}^{\dag } &=&\cos \left( \frac{\theta _{B}}{2}\right) c_{0,X}^{\dag
}+\sin \left( \frac{\theta _{B}}{2}\right) c_{1,X}^{\dag },  \label{c1} \\
b_{AB,K,X}^{\dag } &=&\sin \left( \frac{\theta _{B}}{2}\right) c_{0,X}^{\dag
}-\cos \left( \frac{\theta _{B}}{2}\right) c_{1,X}^{\dag }.  \label{c2}
\end{eqnarray}%
Below we use the convention that pseudospin up corresponds to state $%
\left\vert B,K\right\rangle $\ and pseudospin down to state $\left\vert
AB,K\right\rangle $ and we denote the pseudospin by $\mathbf{S}.$

In our GRPA system of equations for the collective modes, the
orbital-pseudospin wave mode is decoupled from all other modes. Below we
follow one possible strategy for explaining the physics of this mode by
comparing our microscopic GRPA equation of motion to the equations of motion
of an effective orbital pseudospin model, and using the comparison to
identify the effective pseudospin interactions. Since collective modes
correspond to small oscillations of the pseudospin around its quantization
direction, we can use an effective model which has interactions only between
transverse spins. (Quantization direction interactions can be represented as
transverse interactions because of the spin-magnitude constraint.) We write
the pseudospin effective Hamiltonian in momentum ($\mathbf{q}$) space: 
\begin{equation}
H=N_{\varphi }\sum_{\mathbf{q}}J_{ij}\left( \mathbf{q}\right) S_{i}\left( 
\mathbf{q}\right) S_{j}\left( -\mathbf{q}\right) ,  \label{pseudospinH}
\end{equation}%
where $i,j=x,y.$ Since $J_{ij}\left( \mathbf{r},\mathbf{r}^{\prime }\right) $
in the real space version of Eq. (\ref{pseudospinH}) depends on $\mathbf{r}-%
\mathbf{r}^{\prime }$ only, it follows that we can always write $J_{ij}(%
\mathbf{r})=J_{ji}(-\mathbf{r})$ and hence $J_{ij}\left( \mathbf{q}\right)
=J_{ji}\left( -\mathbf{q}\right) $. Because the real-space interactions must
be real we also have the usual property that $J_{ij}\left( \mathbf{q}\right)
=J_{ij}^{\ast }\left( -\mathbf{q}\right) $. Combining these two identities
we can conclude that $J_{xx}\left( \mathbf{q}\right) $ and $J_{yy}\left( 
\mathbf{q}\right) $ are real and even in $\mathbf{q}$, while $J_{xy}\left( 
\mathbf{q}\right) $ has even real and odd imaginary contributions. The real
parts of $J_{xy}\left( \mathbf{q}\right) $ and $J_{yx}\left( \mathbf{q}%
\right) $ are identical while their imaginary parts differ in sign. As we
emphasize further below, the DM interaction is captured by the imaginary
part of $J_{xy}\left( \mathbf{q}\right) $.

Using these properties and the commutation relation,%
\begin{equation}
N_{\varphi }\left[ S_{x}\left( \mathbf{q}\right) ,S_{y}\left( \mathbf{q}%
^{\prime }\right) \right] =\frac{1}{2}i\delta _{\mathbf{q},-\mathbf{q}%
^{\prime }},
\end{equation}%
the equations of motion of the pseudospin model are: 
\begin{equation}
\left( 
\begin{array}{cc}
-iJ_{xy}-\omega & -iJ_{yy} \\ 
iJ_{xx} & iJ_{yx}-\omega%
\end{array}%
\right) \left( 
\begin{array}{c}
S_{x} \\ 
S_{y}%
\end{array}%
\right) =\left( 
\begin{array}{c}
0 \\ 
0%
\end{array}%
\right)  \label{Jdisper}
\end{equation}%
with the dispersion relations%
\begin{equation}
2\omega _{\pm }=-i\left( J_{xy}-J_{yx}\right) \pm \sqrt{4J_{xx}J_{yy}-\left(
J_{xy}+J_{yx}\right) ^{2}}.  \label{disper}
\end{equation}%
Note that the first term on the right hand side of Eq. (\ref{disper}) is
real and odd in $\mathbf{q}$ and that it represents the contribution of the
DM interaction to the collective mode frequency.

Comparing with our microscopic GRPA results for the equations of motion and
collective mode frequencies, we obtain the following expressions for the
pseudospin effective interactions: 
\begin{eqnarray}
J_{xx} &=&\frac{1}{2}A_{3}+\Re \left[ A_{2}\right] -\sin ^{2}(\theta
_{B})\left( \frac{1}{2}A_{1}+\Re \left[ A_{2}\right] \right)  \label{integ}
\\
J_{yy} &=&\frac{1}{2}A_{3}-\Re \left[ A_{2}\right] ,  \notag \\
J_{xy} &=&-\frac{i}{2}\allowbreak \sin (\theta _{B})A_{4}-\cos (\theta
_{B})\Im \left[ A_{2}\right] ,  \notag \\
J_{yx} &=&\frac{i}{2}\allowbreak \sin (\theta _{B})A_{4}-\cos (\theta
_{B})\Im \left[ A_{2}\right] ,  \notag
\end{eqnarray}%
with 
\begin{eqnarray}
A_{1}(q) &=&h_{1}(q)-x_{1}(q)+x_{16}(q)-h_{16}(q),  \label{inter} \\
A_{2}(\mathbf{q}) &=&e^{2i\theta _{\mathbf{q}}}\left( \widehat{h}_{6}\left(
q\right) -\widehat{x}_{6}\left( q\right) \right) ,  \notag \\
A_{3}(q) &=&2h_{1}(q)-2h_{4}(q)-2x_{4}(q)+x_{1}\left( 0\right) ,  \notag \\
A_{4}(\mathbf{q}) &=&-2\Re \left[ ie^{i\theta _{\mathbf{q}}}\left( \widehat{h%
}_{2}\left( q\right) +\widehat{h}_{8}\left( q\right) +\widehat{x}_{2}\left(
q\right) -\widehat{x}_{8}\left( q\right) \right) \right] ,  \notag
\end{eqnarray}%
where $\theta _{\mathbf{q}}$ is the angle between the wavevector $\mathbf{q}$
and the $x$ axis. All interactions are defined in Appendix A. All Hartree
and Fock interaction terms in Eqs. (\ref{inter}), $h_{i},\widehat{h}_{i}$
and $x_{i},\widehat{x}_{i}$, are real and depend only on the modulus of $%
\mathbf{q}.$

We see from the structure of Eqs. (\ref{integ}) that the dispersion relation
has the symmetry $\omega _{\pm }\left( \Delta _{B}^{\left( 2\right) }-\Delta
_{B}\right) =\omega _{\pm }\left( \Delta _{B}\right) .$ Because $\theta
_{B}\left( \Delta _{B}^{\left( 2\right) }-\Delta _{B}\right) =\pi -\theta
_{B}\left( \Delta _{B}\right) $ (from Eq. (\ref{angle})), it follows that
the dispersion at small bias is the same as the dispersion near the critical
bias as first pointed out in Ref. \onlinecite{shizuya1}.

The physical content of the various terms in Eqs. (\ref{integ}) is most
easily identified from their long-wavelength forms. From Eqs. (\ref{integ})
we find that at small $q$ and small $\sin (\theta _{B})\approx \theta
_{B}\approx 2\sqrt{\Delta _{B}/\Delta _{B}^{\left( 2\right) }}$ :

\begin{eqnarray}
J_{xx}\left( \mathbf{q}\right) &\approx &2\beta \Delta _{B}+q\ell \cos
^{2}\left( \theta _{\mathbf{q}}\right)  \label{jxx} \\
&&+\frac{\sqrt{2\pi }}{32}q^{2}\ell ^{2}\left( 1-6\allowbreak \cos
^{2}\left( \theta _{\mathbf{q}}\right) \right) ,  \notag \\
J_{yy}\left( \mathbf{q}\right) &\approx &q\ell \sin ^{2}\left( \theta _{%
\mathbf{q}}\right) \\
&&-\frac{\sqrt{2\pi }}{32}q^{2}\ell ^{2}\left( 5-6\cos ^{2}\left( \theta _{%
\mathbf{q}}\right) \right) \allowbreak ,  \notag \\
J_{xy}\left( \mathbf{q}\right) &\approx &-\frac{1}{\sqrt{2}}i\sqrt{\sqrt{%
\frac{\pi }{2}}\beta \Delta _{B}}\;q\ell \sin \left( \theta _{\mathbf{q}%
}\right) \\
&&-q\ell \sin \left( \theta _{\mathbf{q}}\right) \cos \left( \theta _{%
\mathbf{q}}\right) ,  \notag \\
J_{yx}\left( \mathbf{q}\right) &\approx &\frac{1}{\sqrt{2}}i\sqrt{\sqrt{%
\frac{\pi }{2}}\beta \Delta _{B}}\;q\ell \sin \left( \theta _{\mathbf{q}%
}\right) \\
&&-q\ell \sin \left( \theta _{\mathbf{q}}\right) \cos \left( \theta _{%
\mathbf{q}}\right) .  \notag
\end{eqnarray}%
The pseudospin rotations which change $S_{x}$ correspond to changes in the
angle $\theta $ on the orbital pseudospin Bloch sphere relative to the
ground state value $\theta _{B}$ as illustrated on Fig. \ref{blochs}. For
this reason $J_{xx}$ remains finite (is massive) as $q\rightarrow 0$ when
the potential bias is finite unlike the other couplings. The terms
proportional to $q\ell \cos ^{2}\left( \theta _{\mathbf{q}}\right) $ in $%
J_{xx}$, $q\ell \sin ^{2}\left( \theta _{\mathbf{q}}\right) $ in $J_{yy}$
and $q\ell \sin \left( \theta _{\mathbf{q}}\right) \cos \left( \theta _{%
\mathbf{q}}\right) $ in $J_{xy}$ and $J_{yx}$ are simply electrostatic
interactions between changes generated when the dipole orientation varies in
space. (Recall that the charge density is equal to the divergence of the
dipole density.) These terms are the long-wavelength limits of the Hartree
interactions captured by the GRPA theory. The imaginary contribution to $%
J_{xy}$ is the DM interaction whose physics we discuss below. The
eigenvector for the pseudospin motion, at small $q$ and small $\Delta _{B}$,
has $S_{x}/S_{y}=i\sqrt{2q\ell /\beta \Delta _{B}}\sin \left( \theta _{%
\mathbf{q}}\right) $ if $\theta _{\mathbf{q}}\neq 0$ and $S_{x}/S_{y}=iq\ell 
\sqrt{\sqrt{2\pi }/(16\beta \Delta _{B})}$ if $\theta _{\mathbf{q}}=0$ so
that the long wavelength collective modes are elliptical precessions with
minor axis along the massive $\hat{x}$ direction and major axis along the $%
\hat{y}$ direction which contributes dipolar electrostatic energy. The long
wavelength Goldstone collective mode energy therefore has unusual square
root dispersion: 
\begin{equation}
\omega \left( \mathbf{q}\right) =\sqrt{2\beta \Delta _{B}q\ell }\;\sin
(\theta _{\mathbf{q}}).
\end{equation}%
For $\sin (\theta _{\mathbf{q}})=0$, we have the linear dispersion: 
\begin{equation}
\omega \left( \mathbf{q}\right) =\frac{1}{4}\sqrt{\sqrt{2\pi }\beta \Delta
_{B}}\;q\ell .
\end{equation}%
We see later that the DM interaction assumes a larger importance at larger
bias potentials and shorter wavelengths.

The orbital coherent state occurs at finite bias and is preempted at small
biases by the interlayer coherent state. It is nevertheless interesting to
examine the artificial limit in which $\Delta _{B}\rightarrow 0$, but layer
degrees of freedom are still not in play. In that limit all electrons would
be in the $n=0$ orbital, there would be no electric dipoles in the ground
state, and the exchange parameters would be given by%
\begin{eqnarray}
J_{xx} &=&\frac{1}{2}A_{3}+\Re \left[ A_{2}\right] \\
J_{yy} &=&\frac{1}{2}A_{3}-\Re \left[ A_{2}\right] ,  \notag \\
J_{xy} &=&-\Im \left[ A_{2}\right] ,  \notag \\
J_{yx} &=&-\Im \left[ A_{2}\right] .  \notag
\end{eqnarray}%
The dispersion relation would then be given by 
\begin{equation}
\omega \left( q\right) =\frac{1}{2}\sqrt{A_{3}^{2}\left( q\right)
-4\left\vert A_{2}\left( \mathbf{q}\right) \right\vert ^{2}},
\end{equation}%
which is isotropic. The long-wavelength limit dispersion would become 
\begin{equation}
\omega _{\pm }\left( q\right) \approx \frac{1}{4}\left( \frac{\pi }{2}%
\right) ^{1/4}\left( q\ell \right) ^{3/2},
\end{equation}%
similar to $\nu =-3$ behavior\cite{BarlasPRL}.

\subsection{Moriya interaction and spiral state instability}

As explained previously the DM interaction is captured by the imaginary part
of $J_{xy}(\mathbf{q})$. When this contribution to the orbital pseudospin
Hamiltonian is isolated it yields an interaction of the standard\cite{DM} DM
form: 
\begin{equation}
H_{DM}=iN_{\varphi }\sum_{\mathbf{q}}\Im \left[ J_{xy}\left( \mathbf{q}%
\right) \right] \left( \mathbf{S}\left( \mathbf{q}\right) \times \mathbf{S}%
\left( -\mathbf{q}\right) \right) \cdot \widehat{\mathbf{z}}.
\end{equation}%
An examination of Eqs. (\ref{inter}) shows that this interaction is not due
to electrostatic dipole interactions, and instead to the exchange vertex
corrections i.e. to the interactions $x_{2}\left( q\right) $ and $%
x_{8}\left( q\right) $. In Appendix C, we analyze the exchange energy of
quantum Hall ferromagnets quite generally and show that DM interactions are
the rule rather than the exception when the two-states from which the
pseudospin is constructed have the same spin. The physics of exchange
interaction contributions to collective mode energies is most simply
described by making a particle-hole transformation for occupied states, as
discussed in Appendix C. The exchange interaction at momentum $p$ can then
be related\cite{KallinHalperin} to the attractive interaction between an
electron and a hole separated by $p\ell ^{2}$. DM interactions occur when
the pseudospin state of the electron or hole give rise to cyclotron orbit
charge distributions which do not have inversion symmetry, a property that
holds here because of dipole formation. These distortions of the cyclotron
orbit are irrelevant when $p$ is very large but become important for $p\sim
\ell ^{-1}$. In the case of a simple parabolic band $\nu =1$ quantum Hall
ferromagnet, for example, this picture\cite{KallinHalperin} provides a
simple understanding of the full spin-wave dispersion.

The DM interaction is strongest at $\Delta _{B}/\Delta _{B}^{(2)}=1/2$, 
\emph{i.e.} when $\sin (\theta _{B})=1$. We have found that over a broad
range of $\Delta _{B}$ values the DM interaction is strong enough to induce
an instability of the uniform coherent state. When viewed as a classical
complex-variable quadratic form for Gaussian energy fluctuations, the
orbital pseudospin Hamiltonian, Eq. (\ref{pseudospinH}), is positive
definite provided that $J_{xx}(\mathbf{q})$ is positive, $J_{yy}(\mathbf{q})$
is positive, and 
\begin{equation}
J_{xx}(\mathbf{q})J_{yy}(\mathbf{q})>|J_{xy}(\mathbf{q})|^{2}.  \label{condi}
\end{equation}%
Explicit numerical calculations show that the first two stability
requirements are always satisfied, but that because of the DM interaction,
the third is not satisfied when $\sin (\theta _{B})$ is large. In the GRPA
we find that the OPM first becomes soft at $q_{y}\ell \approx 2$ when $%
\Delta _{B}=\Delta _{B}^{(DM)}\approx $ $\Delta _{B}^{(2)}/10$. (We have $%
\theta _{B}\approx 41^{\circ }$ at this value of $\Delta _{B}$.) Fig. \ref%
{figDM} shows the instability of the orbital-pseudospin mode at $\Delta
_{B}^{(DM)}.$ We remark that the instability occurs at a positive (negative)
value of $q_{y}$ in $\omega _{+}\left( \omega _{-}\right) .$ The
higher-energy collective modes (not shown in the figure) show no sign of
instability.

\begin{figure}[tbph]
\includegraphics[scale=1]{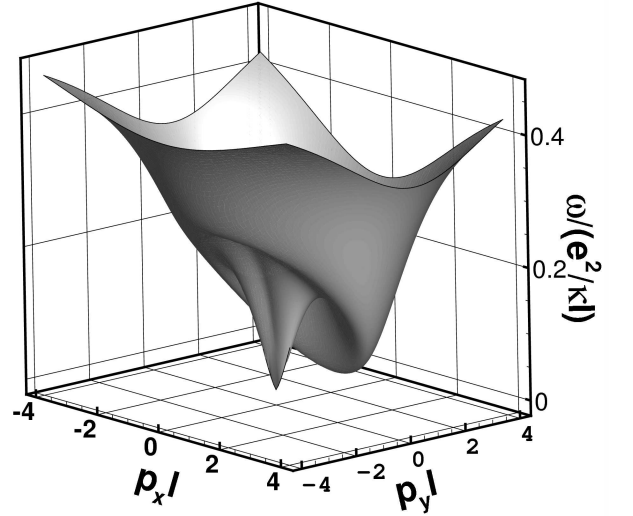}
\caption{Dispersion relation in the orbital-coherent state at the
Dzyaloshinskii-Moriya instability. Filling factor $\protect\nu =-1$ and bias 
$\Delta _{B}=0.51$ $e^{2}/\protect\kappa \ell .$ }
\label{figDM}
\end{figure}

The eigenvector with positive frequency of Eq. (\ref{Jdisper}) has%
\begin{equation}
\frac{S_{x}}{S_{y}}=\frac{-J_{xy}-J_{yx}+i\sqrt{4J_{xx}J_{yy}-\left(
J_{xy}+J_{yx}\right) ^{2}}}{2J_{xx}}.
\end{equation}%
It follows, using Eq. (\ref{condi}), that, at the DM instability, the energy
is lowered by forming coupled density-waves in $\hat{x}$ and $\hat{y}$
pseudospin components with: 
\begin{equation}
\frac{S_{x}(\mathbf{q})}{S_{y}(\mathbf{q})}=-\frac{J_{xy}(\mathbf{q})}{%
J_{xx}(\mathbf{q})}=-\frac{J_{yy}(\mathbf{q})}{J_{xy}^{\ast }(\mathbf{q})}.
\end{equation}

The real part of the $J_{xy}$ coupling, due mainly to the dipole
electrostatic energy, is very small at the instability wavevector because
the $N=0$ Landau level cannot support rapid spatial variation, as we have
verified by explicit calculation. Because $J_{yy}(\mathbf{q})$ is real, it
follows that $S_{x}$ and $S_{y}$ spatial variations are out of phase by
nearly exactly $\pi /2$. If the magnitudes of the $S_{x}$ and $S_{y}$
components were identical, this would imply a spiral ground state. Because $%
J_{xx}(\mathbf{q})$ and $J_{yy}(\mathbf{q})$ are not identical at the
instability, the spiral is somewhat distorted. It must be kept in mind,
however, that the DM\ instability may be preempted by a first order
transition to a state with lower energy and a more complex pseudospin
pattern. A fuller exploration of the properties of these states, including
their properties in the presence of an external electric field, is beyond
the scope of the present work.

\section{COLLECTIVE MODES IN THE MIXED STATE}

The mixed state occurs between the inter-layer coherent and inter-orbital
coherent phases as shown in Fig. \ref{biasnum1}. The width of this region in
the phase diagram increases with magnetic field. In this phase, all order
parameters $\left\langle \rho _{i,j}\right\rangle $ are finite so that this
phase has both inter-layer and inter-orbital coherence. We show the behavior
of some of these order parameters with bias in Fig. \ref{pordre}. The order
parameters $\left\langle \rho _{1,4}\right\rangle $ and $\left\langle \rho
_{2,3}\right\rangle $ which flip both valley and orbital indices are non
zero in this phase.

\begin{figure}[tbph]
\includegraphics[scale=1]{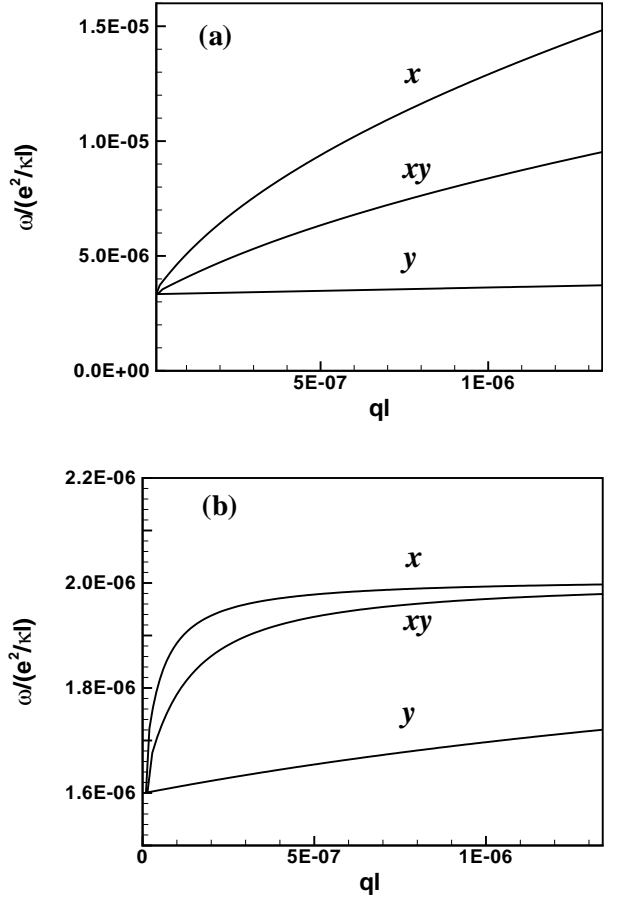}
\caption{ Dispersion relation of the (a) inter-orbital and (b) interlayer
coherent modes in the mixed phase at $\Delta _{B}=0.00195$ $e^{2}/\protect%
\kappa \ell $ and $B=10$ T along directions $x,y$ and $xy$ ($45^{\circ }$
from the $x$ axis). }
\label{mixed}
\end{figure}

The collective modes in the MS are obtained numerically by solving Eq. (\ref%
{grpamodes}). The MS has three dispersive modes, as in the other two phases
we studied. The dispersions of these modes differ from the dispersions in
the other two phases at small wavevector only. We show in Fig. \ref{mixed}%
(a) the dispersion of the inter-orbital mode and in Fig. \ref{mixed}(b) the
dispersion of the inter-layer coherent mode along the $x$ or $y$ axis and at 
$45^{\circ }$ ($xy$) from the $x$ axis. In contrast with the other two
phases, both modes are now gapped for $\Delta _{B}$ not at the boundaries of
this phase. As in the inter-orbital phase, the dispersions are highly
anisotropic. We show in the next section that this phase has a distinct
signature in the microwave absorption spectrum.

\section{MICROWAVE ABSORPTION}

The collective modes discussed in the previous sections can be detected in
microwave absorption experiments, as we now show. We write the current
operator, projected onto $N=0$ and valley $K$, as 
\begin{equation}
j_{\xi K,i}=-c\left. \frac{\partial H_{\xi K}^{0}}{\partial A_{i}^{e}}%
\right\vert _{A_{i}^{e}=0},
\end{equation}%
where $A_{i}^{e}$ is the vector potential of the external electromagnetic
field, $i=x,y$ and $H_{K}^{0}$ is given in Eq. (\ref{deux}). In second
quantization, the total current is given by 
\begin{equation}
\mathbf{J}_{\xi K}=\int d\mathbf{r}\Psi _{\xi K}^{\dag }\left( \mathbf{r}%
\right) \mathbf{j}_{\xi K}\Psi _{\xi K}\left( \mathbf{r}\right) ,
\end{equation}%
with the field operators defined in Eqs. (\ref{field1},\ref{field2}) and $%
\xi =\pm 1.$ We find that 
\begin{equation}
\mathbf{J}_{\xi K}=\xi \sqrt{2}\beta \Delta _{B}\frac{e\ell }{\hslash }%
N_{\varphi }\left( \rho _{\xi K,y}\left( 0\right) \widehat{\mathbf{x}}+\rho
_{\xi K,x}\left( 0\right) \widehat{\mathbf{y}}\right) ,  \label{bb}
\end{equation}%
where 
\begin{eqnarray}
\rho _{K,x}\left( 0\right)  &=&\frac{1}{2}\left( \rho _{0,1}^{K,K}\left(
0\right) +\rho _{1,0}^{K,K}\left( 0\right) \right) , \\
\rho _{K,y}\left( 0\right)  &=&\frac{1}{2i}\left( \rho _{0,1}^{K,K}\left(
0\right) -\rho _{1,0}^{K,K}\left( 0\right) \right) ,
\end{eqnarray}%
and similarly for $\rho _{K^{\prime },x/y}\left( 0\right) .$ The same result
for $\mathbf{J}_{\xi K}$ can be obtained by calculating the polarization
current%
\begin{equation}
\mathbf{J}_{\xi K}=\frac{d}{dt}\mathbf{d}_{\xi K}\left( \mathbf{q}\right) ,
\end{equation}%
with the dipole density $\mathbf{d}_{\xi K}\left( \mathbf{q}\right) $
defined in Eq. (\ref{dipoleq}) and using the Heisenberg equation of motion $%
-i\hslash d/dt\left( \ldots \right) =\left[ H_{\xi K}^{0},\left( \ldots
\right) \right] .$ We note that the orbitally coherent state has spontaneous
currents in its ground state, a very exceptional property. It appears likely
that, in finite systems, these currents should flow perpendicular to system
boundaries, forcing domain structures in the pseudospin magnetization
texture, consistent with expectations based on the electrostatic energy of
the dipole density.

We define the total current-current correlation function Matsubara Green's
functions as 
\begin{eqnarray}
\chi _{J_{\alpha },J_{\beta }}\left( \tau \right) &=&-\frac{1}{S}%
\left\langle TJ_{\alpha }\left( \tau \right) J_{\beta }\left( 0\right)
\right\rangle \\
&=&\left( \frac{\Delta _{B}}{\gamma _{1}}\right) ^{2}\frac{e^{2}\hslash ^{2}%
}{4\pi m^{\ast 2}\ell ^{4}}\sum_{i,j=K,K^{\prime }}\xi _{i}\xi _{j}\chi
_{\rho _{i,\overline{\alpha }},\rho _{j,\overline{\beta }}}\left( \tau
\right) ,  \notag
\end{eqnarray}%
where $J_{\alpha }=J_{K,\alpha }+J_{K^{\prime },\alpha }$ and 
\begin{equation}
\chi _{\rho _{i,\alpha },\rho _{j,\beta }}\left( \tau \right) =-\left\langle
T\rho _{i,\alpha }\left( 0,\tau \right) \rho _{j,\beta }\left( 0,0\right)
\right\rangle .
\end{equation}%
with $i,j=K,K^{\prime }$ and $\alpha ,\beta =x,y.$ Note that $\overline{%
\alpha },\overline{\beta }$ are defined so that $\overline{x}=y,\overline{y}%
=x.$

The microwave absorption for an electric field oriented along the direction $%
\alpha $ is given by%
\begin{eqnarray}
P_{\alpha }\left( \omega \right) &=&-\frac{1}{\hslash }\Im \left[ \frac{\chi
_{J_{\alpha },J_{\alpha }}^{R}\left( \omega \right) }{\omega +i\delta }%
\right] E_{0}^{2}  \label{mwabsorption} \\
&=&-\frac{1}{2}\left( \frac{e^{2}}{h}\right) \left( \frac{\Delta _{B}}{%
\gamma _{1}}\right) ^{2}\omega _{c}^{\ast 2}  \notag \\
&&\times \sum_{i,j=K,K^{\prime }}\Im \left[ \xi _{i}\xi _{j}\frac{\chi
_{\rho _{i,\overline{\alpha },j,\overline{\alpha }}}^{R}\left( \omega
\right) }{\omega +i\delta }\right] E_{0}^{2},  \notag
\end{eqnarray}%
where we have assumed a uniform electric field $\mathbf{E}\left( t\right)
=E_{0}\widehat{\mathbf{\alpha }}e^{-i\omega t}$ and taken the analytic
continuation $i\Omega _{n}\rightarrow \omega +i\delta $ of $\chi _{J_{\alpha
},J_{\alpha }}\left( 0,\Omega _{n}\right) $ to get the retarded response
function. The response functions $\chi _{\rho _{i\alpha }\rho _{j\alpha
}}^{R}$ are calculated in units of $\hslash /\left( e^{2}/\kappa \ell
\right) $ so that $P_{\alpha }\left( \omega \right) $ is the power absorbed
per unit area. In Eq. (\ref{mwabsorption}) we have neglected a diamagnetic
contribution to the current response which becomes important at low
frequencies.

\begin{figure}[tbph]
\includegraphics[scale=1]{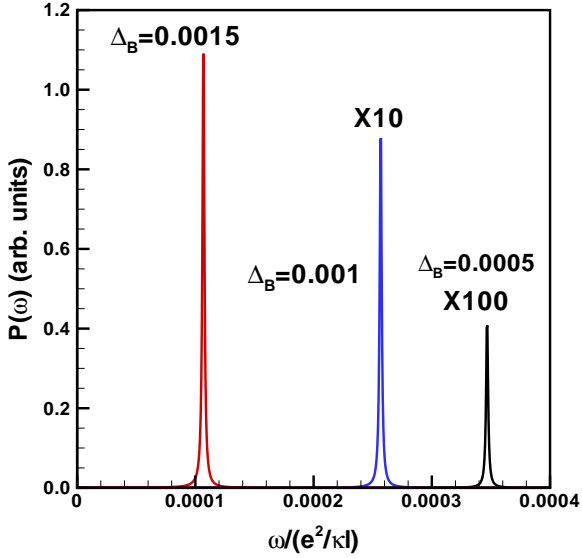}
\caption{ (Color online) Microwave absorption from the orbital pseudospin
mode in the inter-layer phase at different values of the bias. The
absorption is zero at zero bias. The second(third) peak has been multiplied
by 10(100). }
\label{abso}
\end{figure}

Our GRPA correlation functions are given by Eq. (\ref{grpamodes}) and
numerical results for the absorption in the inter-layer coherent phase are
shown in Fig. \ref{abso}. Exactly the same result is obtained, in this
phase, if the electric field is set in the $y$ direction, \emph{i.e.} the
absorption is isotropic. We see that the signal in the absorption is at a
frequency corresponding to the orbital-pseudospin mode (see Fig. \ref%
{modesnum1petitq}). The frequency $\nu $ of this mode at $\mathbf{q}=0$
decreases with bias while the absorption intensity increases with bias.
Since $e^{2}/\hbar \kappa \ell \sim 2.7\times 10^{3}\sqrt{B\left( \text{T}%
\right) }$ GHz (using $\kappa =5$ for graphene on SiO$_{2}$ substrate), the
frequency of the orbital pseudospin mode at zero bias is $\nu \approx 3.4$
GHz $i.e.$ in the microwave regime. Note that mode $3$ in Fig. \ref%
{modesnum1} is also present in the absorption at finite bias and that its
frequency has a higher value outside of the microwave regime.

\begin{figure}[tbph]
\includegraphics[scale=1]{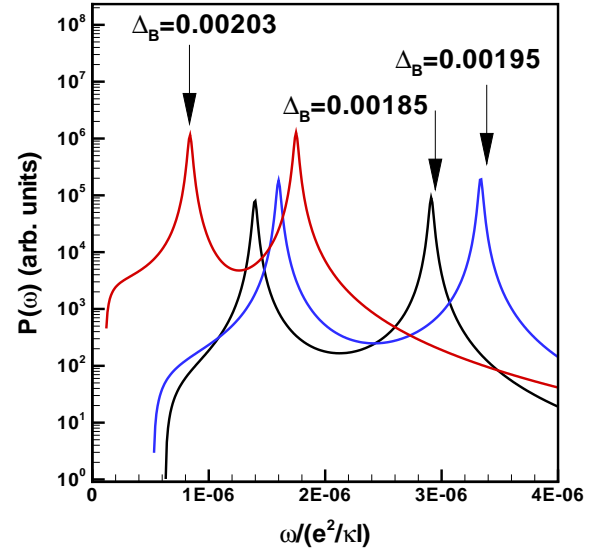}
\caption{ (Color online) Absorption spectrum in the mixed state at $B=10$ T
for an electric field in the $y$ direction and for biases $\Delta
_{B}=0.00185,0.00195,0.0020$ in units of $e^{2}/\protect\kappa \ell $ }
\label{absomixed}
\end{figure}

The absorption in the mixed state is shown in Fig. \ref{absomixed} on a
logarithmic scale. The three lower peaks show the absorption from the
inter-layer coherent mode which is gapped in the mixed state. The other
three peaks are from the inter-orbital mode. The gaps in these two modes
increase with bias until they reach a maximum around $\Delta _{B}=0.00195$
for $B=10$ T. The gaps then decreases with bias. The intensity of the
absorption increases with bias for both modes. In Fig. \ref{absomixed}, the
electric field is set along the $y$ axis. The absorption is at least $100$
times lower if the electric field is oriented along the $x$ axis i.e. it is
highly anisotropic in the mixed state. With our choice of phase for the
ground state of the mixed state, the pseudospins (and so the electric
dipoles according to Eq. (\ref{dipoled})) are oriented in the $\widehat{%
\mathbf{y}}$ direction. Their motion for $q=0$ is an oscillation in the $x-y$
plane about the $y$ axis. For such a configuration, the electromagnetic
absorption is strongest for fields along the $y$ axis, which is what we
observe in our calculation.

In the orbital-coherent phase, the orbital-pseudospin mode is gapless and
decoupled from the two other gapped modes. Since the orbital-coherent mode
couples strongly to external electric fields, we can expect anomalous
low-frequency absorption in this state, similar to the Drude absorption of a
metal. This interesting and unusual absorption feature is likely to be
highly sensitive to disorder. Its detailed analysis lies beyond the scope of
the present paper. %
Above $\Delta _{B}^{\left( 2\right) }$, the ground state has $\nu _{2}=\nu
_{3}=\nu _{4}=1$ and there is no orbital coherence anymore. The OPM then has
a gap that is proportional to $\Delta _{B}-\Delta _{B}^{\left( 2\right) }.$
The OPM becomes visible in the absorption in this phase while the other
modes do not.

\begin{figure}[tbph]
\includegraphics[scale=1]{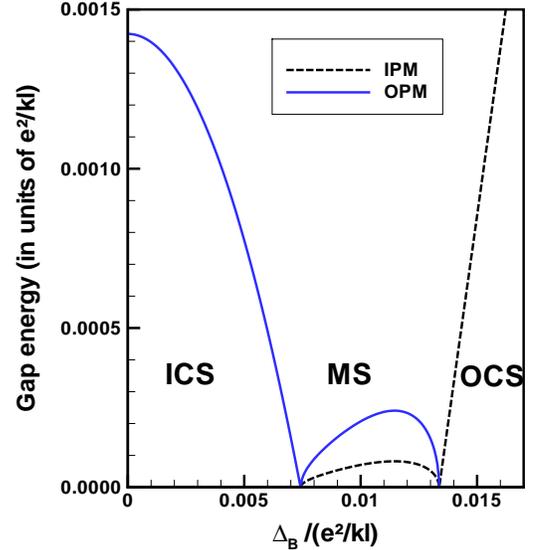}
\caption{ (Color online) Gaps in the inter-layer pseudospin mode (IPM)\ and
in the orbital pseudospin mode (OPM) as a function of bias at $\protect\nu %
=-1$ and $B=40$ T. }
\label{abso40}
\end{figure}

In summary, we see that each of the three phases in the phase diagram at $%
\nu =-1$ has a different signature in the microwave absorption spectrum. We
remark that the frequencies in Fig. \ref{absomixed} are quite small. But,
they can be increased by increasing the magnetic field. For example, Fig. %
\ref{abso40} shows the gap in the inter-layer and inter-orbital pseudospin
modes in the three phases at a magnetic field of $40$ T.

\section{CONCLUSION}

We have studied the phase diagram and collective excitations of a
spin-polarized bilayer graphene 2DEG at filling factors $\nu =-1$ and $\nu
=3 $, as a function of a bias electric potential which shifts electrons
between layers. Our study is based on the Hartree-Fock approximation for the
mean-field ground state, and the GRPA\ for the collective modes and response
functions.

We predict phase transitions between the following sequence of states with
increasing potential bias: (1) an inter-layer-coherent state (ICS) with a
zero gap inter-layer pseudospin mode (IPM)\ and an orbital pseudospin mode
(OPM)\ with a small gap. This gap can be detected in microwave absorption
experiment. Its frequency decreases with bias while its intensity increases
with bias; (2) a mixed state (MS) with both inter-layer and inter-orbital
coherence. Both the IPM and OPM are gapped and visible in microwaves in this
phase. Moreover, the intensity of the absorption is highly sensitive to the
direction of the external electric field; (3) an orbital coherent state
(OCS) with orbital coherence concentrated in one layer only. (The second
layer is completely filled.) The OCS state is a very simple one in which the
low potential layer is filled and the high potential layer has a gap induced
between it's two Landau levels by spontaneously establishing coherence
between states with $n=1$ and $n=0$ orbital character. This state has a
number of quite unusual properties, including electric dipoles and
associated uniform currents in its ground state. The OCS has a gapless
(Goldstone) OPM and a gapped IPM. Both modes are absent from the absorption
spectrum. The dispersions of the collective modes are also highly
anisotropic in this phase. The phase is unstable at a finite wavevector due
to the presence of a Dzyaloshinksii-Moriya exchange interaction. We believe
that the instability will lead to the formation of a ground state with a
non-uniform pseudosopin pattern.

These properties are associated with competition between an
electron-electron interaction term which favors $n=0$ orbital occupation,
and single-particle terms in the effective two-band model of Ref. %
\onlinecite{mccann} which favors $n=1$ orbital occupation more strongly at
larger inter-layer potential difference. We note that this band model
neglect the $\gamma _{4}$ hopping term, which connects sites $A_{1}(B_{1})$
and $A_{2}(B_{2})$\cite{graphenereviews}. Strictly speaking, the results
obtained in our paper are valid if $\gamma _{4}=0.$ The effect of the $%
\gamma _{4}$ term is to add a negative correction to the energy of the $n=1$
orbital which is independent of the valley index, magnetic field and bias.
More precisely: $E_{\mathbf{K,}n=1,X}=\frac{1}{2}\Delta _{B}-\beta \Delta
_{B}+\zeta ,E_{\mathbf{K}^{\prime },n=1,X}=-\frac{1}{2}\Delta _{B}+\beta
\Delta _{B}+\zeta $ with $\zeta =\sqrt{6\beta }\gamma _{4}a_{0}/\ell .$ It
is not clear from the literature what is the precise value of the $\gamma
_{4}$ parameter. In Ref. \onlinecite{peeters}, the value $\gamma _{4}=-0.12$
is obtained by comparing the tight-binding model for bilayer graphene with
the Slonczewski-Weiss-McClure tight-binding model for bulk graphite\cite%
{mcclure}. In bulk graphite, $\gamma _{4}\approx 0.044$ eV. With $\gamma
_{4}=-0.12$, the correction $\zeta <\Delta _{B}$ for $\Delta _{B}>0.02$ $%
e^{2}/\kappa \ell .$ It follows that this term may have an effect on the
existence of the interlayer coherent and mixed phases, and certainly on the
location of their phase boundaries, $\Delta _{B,num}^{(3)}$ and $\Delta
_{B}^{(1)}$, since level $n=1$ may be below level $n=0$ in \textit{both}
layers. The Dzyaloshinksii-Moriya transition, however, occurs at a large
bias of $\Delta _{B}\approx 0.5$ $e^{2}/\kappa \ell $ in a region where the
single-particle energy of level $n=1$ is already well below that of level $%
n=0.$ For this reason we believe that the physics in this region should not
be affected by $\gamma _{4}.$ More numerical work is needed, however, to
assess the precise influence of this term.

\begin{acknowledgments}
A.-H. MacDonald was supported by the NSF\ under grant DMR-0606489 and by the
Welch Foundation. R. C\^{o}t\'{e} was supported by a grant from the Natural
Sciences and Engineering Research Council of Canada (NSERC). Y. Barlas was
supported by a grant from the State of Florida. Computer time was provided
by the R\'{e}seau Qu\'{e}b\'{e}cois de Calcul Haute Performance (RQCHP).
\end{acknowledgments}

\appendix{}

\section{HARTREE AND FOCK INTERACTIONS}

We first give the definitions of the Hartree, $H,$ and Fock, $X,$
interactions in Eqs. (\ref{hart1},\ref{fock1},\ref{hart2},\ref{fock2}): 
\begin{eqnarray}
h_{1}\left( q\right) &=&H_{0,0,0,0}\left( q\right) =\frac{1}{q\ell }\Lambda
\left( q\right) , \\
h_{2}\left( \mathbf{q}\right) &=&H_{0,0,0,1}\left( \mathbf{q}\right) =-\frac{%
i}{\sqrt{2}}e^{i\theta _{\mathbf{q}}}\Lambda \left( q\right) , \\
h_{4}\left( q\right) &=&H_{0,0,1,1}\left( q\right) =\frac{1}{q\ell }\left( 1-%
\frac{q^{2}\ell ^{2}}{2}\right) \Lambda \left( q\right) , \\
h_{6}\left( \mathbf{q}\right) &=&H_{0,1,0,1}\left( \mathbf{q}\right) =\frac{1%
}{2}q\ell e^{2i\theta _{\mathbf{q}}}\Lambda \left( q\right) , \\
h_{7}\left( q\right) &=&H_{0,1,1,0}\left( q\right) =\frac{1}{2}q\ell \Lambda
\left( q\right) , \\
h_{8}\left( \mathbf{q}\right) &=&H_{0,1,1,1}\left( \mathbf{q}\right) =\frac{i%
}{\sqrt{2}}e^{i\theta _{\mathbf{q}}} \\
&&\times \left( 1-\frac{q^{2}\ell ^{2}}{2}\right) \Lambda \left( q\right) , 
\notag \\
h_{16}\left( q\right) &=&H_{1,1,1,1}\left( q\right) =\frac{1}{q\ell }\left(
1-\frac{q^{2}\ell ^{2}}{2}\right) \Lambda \left( q\right)
\end{eqnarray}%
and {%
\begin{eqnarray}
h_{3} &=&H_{0,0,1,0}\left( \mathbf{q}\right) =-h_{2}^{\ast }, \\
h_{5} &=&H_{0,1,0,0}\left( \mathbf{q}\right) =-h_{2}, \\
h_{9} &=&H_{1,0,0,0}\left( \mathbf{q}\right) =h_{2}^{\ast }, \\
h_{10} &=&H_{1,0,0,1}\left( q\right) =h_{7}, \\
h_{11} &=&H_{1,0,1,0}\left( \mathbf{q}\right) =h_{6}^{\ast },
\end{eqnarray}%
\begin{eqnarray}
h_{12} &=&H_{1,0,1,1}\left( \mathbf{q}\right) =-h_{8}^{\ast }, \\
h_{13} &=&H_{1,1,0,0}\left( q\right) =h_{4}, \\
h_{14} &=&H_{1,1,0,1}\left( \mathbf{q}\right) =-h_{8}, \\
h_{15} &=&H_{1,1,1,0}\left( \mathbf{q}\right) =h_{8}^{\ast },
\end{eqnarray}%
where }$\theta _{\mathbf{q}}$ is the angle between the wavevector $\mathbf{q}
$ and the $x$ axis and $\Lambda \left( q\right) =\exp \left( \frac{%
-q^{2}\ell ^{2}}{2}\right) .$ The interactions $\widetilde{h}_{n}\left( 
\mathbf{q}\right) $ are obtained by multiplying $h_{n}$ by $e^{-qd}$ where $%
d $ is the inter-layer separation. The interactions $\widehat{h}_{n}\left( 
\mathbf{q}\right) $ and $\widehat{\widetilde{h}}_{n}$ are obtained by
removing the term $i$ and the phase factor $e^{\pm i\theta _{\mathbf{q}}}$
or $e^{\pm 2i\theta _{\mathbf{q}}}$ in $h_{n}$ and $\widetilde{h}_{n}$. For
example $\widehat{h}_{2}=-\Lambda \left( q\right) /\sqrt{2}$ while $%
h_{2}\left( \mathbf{q}\right) =-\frac{i}{\sqrt{2}}e^{i\theta _{\mathbf{q}%
}}\Lambda \left( q\right) .$

The Fock interactions are defined by%
\begin{equation}
x_{1}\left( q\right) =X_{0,0,0,0}\left( \mathbf{q}\right) =\int_{0}^{\infty
}dye^{-y^{2}/2}J_{0}\left( q\ell y\right) ,
\end{equation}%
\begin{equation}
x_{2}\left( \mathbf{q}\right) =X_{0,0,0,1}\left( \mathbf{q}\right) =\frac{i}{%
\sqrt{2}}e^{i\theta _{\mathbf{q}}}\int_{0}^{\infty
}dyye^{-y^{2}/2}J_{1}\left( q\ell y\right) ,
\end{equation}%
\begin{equation}
x_{4}\left( q\right) =X_{0,0,1,1}\left( \mathbf{q}\right) =\int_{0}^{\infty
}dy\left( 1-\frac{y^{2}}{2}\right) e^{-y^{2}/2}J_{0}\left( q\ell y\right) ,
\end{equation}%
\begin{equation}
x_{6}\left( \mathbf{q}\right) =X_{0,1,0,1}\left( \mathbf{q}\right) =\frac{1}{%
2}e^{2i\theta _{\mathbf{q}}}\int_{0}^{\infty }dyy^{2}e^{-y^{2}/2}J_{2}\left(
q\ell y\right) ,
\end{equation}%
\begin{equation}
x_{7}\left( q\right) =X_{0,1,1,0}\left( \mathbf{q}\right) =\frac{1}{2}%
\int_{0}^{\infty }dyy^{2}e^{-y^{2}/2}J_{0}\left( q\ell y\right) ,
\end{equation}%
\begin{eqnarray}
x_{8}\left( \mathbf{q}\right) &=&X_{0,1,1,1}\left( \mathbf{q}\right) =-\frac{%
i}{\sqrt{2}}e^{i\theta _{\mathbf{q}}}\int_{0}^{\infty }dyy \\
&&\times \left( 1-\frac{y^{2}}{2}\right) e^{-y^{2}/2}J_{1}\left( q\ell
y\right) ,  \notag
\end{eqnarray}%
\begin{equation}
x_{16}\left( q\right) =X_{1,1,1,1}\left( \mathbf{q}\right) =\int_{0}^{\infty
}dy\left( 1-\frac{y^{2}}{2}\right) ^{2}e^{-y^{2}/2}J_{0}\left( q\ell
y\right) ,
\end{equation}%
and 
\begin{eqnarray}
x_{3} &=&X_{0,0,1,0}\left( \mathbf{q}\right) =x_{2}^{\ast }, \\
x_{5} &=&X_{0,1,0,0}\left( \mathbf{q}\right) =-x_{2,} \\
x_{9} &=&X_{1,0,0,0}\left( \mathbf{q}\right) =-x_{2}^{\ast }, \\
x_{10} &=&X_{1,0,0,1}\left( q\right) =x_{7}, \\
x_{11} &=&X_{1,0,1,0}\left( \mathbf{q}\right) =x_{6}^{\ast },
\end{eqnarray}%
\begin{eqnarray}
x_{12} &=&X_{1,0,1,1}\left( \mathbf{q}\right) =x_{8}^{\ast }, \\
x_{13} &=&X_{1,1,0,0}\left( q\right) =x_{4}, \\
x_{14} &=&X_{1,1,0,1}\left( \mathbf{q}\right) =-x_{8}, \\
x_{15} &=&X_{1,1,1,0}\left( \mathbf{q}\right) =-x_{8}^{\ast }.
\end{eqnarray}%
The interactions $\widetilde{x}_{n}$ are obtained by multiplying the
integrand by $e^{-yd/\ell }$. The interactions $\widehat{x}_{n}$ and $%
\widehat{\widetilde{x}}_{n}$ are obtained by removing the imaginary term $i$
and the phase factor.

The combinations:%
\begin{eqnarray}
H_{i} &=&h_{i}-\widetilde{h}_{i}, \\
T_{i} &=&h_{i}+\widetilde{h}_{i}, \\
X_{i} &=&x_{i}+\widetilde{x}_{i}, \\
U_{i} &=&x_{i}-\widetilde{x}_{i}.
\end{eqnarray}%
To define $\widehat{H}_{n},\widehat{T}_{n},\widehat{X}_{n},\widehat{U}_{n},$
we follow the same procedure as for $\widehat{h}_{n},\widehat{\widetilde{h}}%
_{n},\widehat{x}_{n},\widehat{\widetilde{x}}_{n}.$

Some useful constants: 
\begin{equation}
x_{1}\left( 0\right) =\sqrt{\frac{\pi }{2}},
\end{equation}%
\begin{equation}
x_{4}\left( 0\right) =\frac{1}{2}\sqrt{\frac{\pi }{2}},
\end{equation}%
\begin{equation}
x_{7}\left( 0\right) =\frac{1}{2}\sqrt{\frac{\pi }{2}},
\end{equation}%
\begin{equation}
x_{16}\left( 0\right) =\frac{3}{4}\sqrt{\frac{\pi }{2}}.
\end{equation}

\section{MATRIX $F_{1}$ FOR THE INTER-LAYER-COHERENT MODES}

The collective modes at $\nu =-1$ and $\Delta _{B}<\Delta _{B}^{(1)}$
involve the matrices $F_{1}\left( q\right) $ in Eq. (\ref{f1q}). The
elements of this matrix are defined by 
\begin{eqnarray}
A\left( q\right) &=&\widetilde{x}_{4}\left( 0\right) -\widetilde{x}%
_{1}\left( 0\right) -\frac{3}{4}x_{1}\left( 0\right) -\widetilde{x}%
_{16}\left( 0\right) \\
&&+X_{4}\left( q\right) -H_{1}\left( q\right) +H_{4}\left( q\right) ,  \notag
\end{eqnarray}%
\begin{eqnarray}
B\left( q\right) &=&2\Delta _{B}^{c}\left( \beta -1\right) +H_{1}\left(
q\right) -H_{4}\left( q\right) -U_{4}\left( q\right) \\
&&+\widetilde{x}_{1}\left( 0\right) -\frac{5}{4}x_{1}\left( 0\right) -%
\widetilde{x}_{4}\left( 0\right) +\widetilde{x}_{16}\left( 0\right) +2\frac{d%
}{\ell },  \notag
\end{eqnarray}%
\begin{equation}
C\left( q\right) =-H_{16}\left( q\right) +X_{16}\left( q\right) -2\widetilde{%
x}_{16}\left( 0\right) ,
\end{equation}%
\begin{eqnarray}
D\left( q\right) &=&2\Delta _{B}^{c}\left( 2\beta -1\right) +H_{16}\left(
q\right) -U_{16}\left( q\right) \\
&&+2\widetilde{x}_{16}\left( 0\right) -\frac{3}{2}x_{1}\left( 0\right) +2%
\frac{d}{\ell },  \notag
\end{eqnarray}%
\begin{eqnarray}
E\left( q\right) &=&T_{4}\left( q\right) +X_{4}\left( q\right) -T_{1}\left(
q\right) +\widetilde{x}_{1}\left( 0\right) \\
&&-\widetilde{x}_{4}\left( 0\right) -\frac{3}{4}x_{1}\left( 0\right) -%
\widetilde{x}_{16}\left( 0\right) ,  \notag
\end{eqnarray}%
\begin{eqnarray}
F\left( q\right) &=&2\beta \Delta _{B}^{c}-H_{1}\left( q\right) +H_{4}\left(
q\right) +U_{4}\left( q\right) \\
&&+\widetilde{x}_{4}\left( 0\right) -\widetilde{x}_{1}\left( 0\right) -\frac{%
1}{4}x_{1}\left( 0\right) +\widetilde{x}_{16}\left( 0\right) ,  \notag
\end{eqnarray}%
\begin{eqnarray}
G\left( q\right) &=&\Delta _{B}^{c}-H_{1}\left( q\right) +H_{4}\left(
q\right) +U_{4}\left( q\right) \\
&&+\widetilde{x}_{4}\left( 0\right) -\widetilde{x}_{1}\left( 0\right) +\frac{%
1}{2}x_{1}\left( 0\right) -\frac{d}{\ell },  \notag
\end{eqnarray}%
\begin{equation}
H\left( q\right) =\widehat{H}_{6}\left( q\right) -\widehat{U}_{6}\left(
q\right) ,
\end{equation}%
\begin{equation}
J\left( q\right) =\widehat{H}_{8}\left( q\right) -\widehat{U}_{8}\left(
q\right) ,
\end{equation}%
\begin{equation}
K\left( q\right) =\widehat{X}_{6}\left( q\right) -\widehat{T}_{6}\left(
q\right) ,
\end{equation}%
\begin{equation}
M\left( q\right) =H_{16}\left( q\right) -U_{16}\left( q\right) ,
\end{equation}%
\begin{equation}
N\left( q\right) =\widehat{H}_{8}\left( q\right) -\widehat{X}_{8}\left(
q\right) .
\end{equation}

To get the gap given by Eq. (\ref{gaporbi}), we use 
\begin{eqnarray}
\left\vert E\left( 0\right) \right\vert &=&\left\vert -\frac{1}{4}\sqrt{%
\frac{\pi }{2}}+\widetilde{x}_{1}\left( 0\right) -\widetilde{x}_{16}\left(
0\right) \right\vert  \notag \\
&=&x_{1}\left( 0\right) -\widetilde{x}_{1}\left( 0\right) -x_{16}\left(
0\right) +\widetilde{x}_{16}\left( 0\right) , \\
K\left( 0\right) &=&0.
\end{eqnarray}

\section{EXCHANGE ENERGY PSEUDOSPIN DEPENDENCE}

It is possible to derive a rather general and instructive expression for the
exchange energy of a pseudospin-$1/2$ quantum Hall ferromagnet, for the case
in which the pseudospin texture varies in one direction only. In the
following we take this direction to be the $\hat{x}$ direction and choose a
Landau gauge in which the guiding centers orbits are localized as a function
of this coordinate. If we are interested only in the dependence of energy on
pseusospin texture we can assume that every guiding center orbital is
occupied by one electron, but leave the pseudospin of that orbital
arbitrary. It is not necessary to immediately specify the orbital character
of the states that form the pseudospin and we refer to them for the moment
as state A and state B. In the case of immediate interest in this paper,
state A has $n=0$ orbital character and state B has $n=1$ orbital character.
We discuss some other examples below.

The pseudospin texture can be specified by the two-component pseudospinors
at each guiding center, 
\begin{equation}
|\Psi _{X}\rangle =\left( 
\begin{array}{c}
z_{X}^{A} \\ 
z_{X}^{B}%
\end{array}%
\right) ,
\end{equation}%
or by the guiding center dependent direction cosines of the pseudospin
orientation: $(n_{X}^{x},n_{X}^{y},n_{X}^{z})=(\sin (\theta _{X})\cos (\phi
_{X}),\sin (\theta _{X})\sin (\phi _{X}),\cos (\theta _{X}))$ where $\theta
_{X}$ and $\phi _{X}$ are the pseudospin orientation polar and azimuthal
angles. When a specific gauge choice is convenient we use $z^{A}=\cos
(\theta /2)$ and $z^{B}=\sin (\theta /2)\exp (i\phi )$.

The exchange energy of a pseudospin quantum Hall ferromagnet is 
\begin{equation}
E_{x}=-\frac{1}{2}\sum_{X,X^{\prime }}\;\langle X,X^{\prime
}|V_{ee}|X^{\prime },X\rangle .
\end{equation}%
The dependence of exchange energy on pseudospin texture can be exhibited
explicitly by using the property that each $\exp (i\mathbf{k\cdot
(r_{1}-r_{2})})$ term in the Fourier expansion of the two-particle
interaction matrix elements can be separated into factors that depend on ${%
\mathbf{r}_{1}}$ and ${\mathbf{r}_{2}}$ independently. It follows that 
\begin{equation}
E_{x}=-\frac{1}{2}\sum_{X,X^{\prime }}\;n_{X^{\prime }}^{\alpha
}\,J_{X^{\prime },X}^{\alpha ,\beta }\,n_{X}^{\beta }  \label{eq:ex}
\end{equation}%
where the Greek indices $\alpha =c,x,y,z$, $n^{c}\equiv 1$ represents the
filling factor of the guiding center states, 
\begin{equation}
J_{X^{\prime },X}^{\alpha ,\beta }=\frac{1}{A}\sum_{\mathbf{q}}\exp
(-q^{2}\ell ^{2}/2)\;f^{\alpha ,\beta }({\mathbf{q}})\;V(\mathbf{q})\;\delta
_{X^{\prime }-X,\ell ^{2}q_{y}},  \label{eq:jab}
\end{equation}%
$\ell $ is the magnetic length, $V(\mathbf{q})$ is the Fourier-transform of
the electron-electron interaction and $f^{\alpha ,\beta }({\mathbf{q}})$ is
an interaction form factor. Eq. (\ref{eq:jab}) follows from the following
expression for the plane-wave matrix elements: 
\begin{eqnarray}
\langle X^{\prime }|\exp (i\mathbf{q\cdot r})|X\rangle &=&\delta _{X^{\prime
},X+\ell ^{2}q_{y}}\exp (-q^{2}\ell ^{2}/4)  \notag  \label{FIPI} \\
&\times &\sum_{I^{\prime },I}\;{\bar{z}}_{X^{\prime }}^{I^{\prime
}}\,F^{I^{\prime },I}({\mathbf{q}})\,z_{X}^{I},
\end{eqnarray}%
where the overbar accent denotes complex conjugation. The character of the
cyclotron orbitals in the two nearly degenerate Landau levels is captured by
the single-particle form factors $F^{I^{\prime },I}({\mathbf{q}})$. For
example if pseudospin $I$ has orbital Landau level index $n_{I}$, $%
F^{I^{\prime },I}({\mathbf{q}})=F_{n_{I^{\prime }},n_{I}}({\mathbf{q}})$
where $F_{n^{\prime },n}({\mathbf{q}})$ is the familiar two-dimensional
electron gas Landau-level form factor\cite{leshouches}, commonly used in the
analysis of many-different properties in the quantum Hall regime. Note that $%
F^{I^{\prime },I}(\mathbf{q})={\bar{F}}^{I,I^{\prime }}(\mathbf{-q})$ For
the example of quantum Hall ferromagnetism discussed in this paper, $n_{A}=0$
and $n_{B}=1$ when the bias potential is strong enough to yield complete
layer polarization. (The analysis in this section does not apply when both
orbital and layer degrees of freedom are in play.) When the $A$ and $B$
orbitals have opposite spins, a common occurrence in quantum Hall
ferromagnetism, $F^{A,B}({\mathbf{q}})\equiv 0$. Examples in which the $A$
and $B$ orbitals are centered in different two-dimensional layers require a
slight generalization of the present discussion which we do not explicitly
address.

Using Eq. (\ref{FIPI}) we find (leaving the wavevector dependence of the
single-particle and interaction form factors implicit) that 
\begin{eqnarray}
4f^{0,0} &=&|F^{A,A}|^{2}+|F^{B,B}|^{2}+|F^{A,B}|^{2}+|F^{B,A}|^{2}  \notag
\\
4f^{0,x} &=&2\mathrm{Re}[\bar{F}^{A,A}F^{A,B}+\bar{F}^{B,A}F^{B,B}]  \notag
\\
4f^{0,y} &=&-2\mathrm{Im}[\bar{F}^{A,A}F^{A,B}+\bar{F}^{B,A}F^{B,B}]  \notag
\\
4f^{0,z} &=&|F^{A,A}|^{2}-|F^{B,B}|^{2}-|F^{A,B}|^{2}+|F^{B,A}|^{2}  \notag
\\
4f^{x,x} &=&2\mathrm{Re}[F^{A,A}\bar{F}^{B,B}+F^{B,A}\bar{F}^{A,B}]  \notag
\\
4f^{y,y} &=&2\mathrm{Re}[F^{A,A}\bar{F}^{B,B}-F^{B,A}\bar{F}^{A,B}]  \notag
\\
4f^{z,z} &=&|F^{A,A}|^{2}+|F^{B,B}|^{2}-|F^{A,B}|^{2}-|F^{B,A}|^{2}  \notag
\\
4f^{x,y} &=&2\mathrm{Im}[F^{A,A}\bar{F}^{B,B}+F^{B,A}\bar{F}^{A,B}]  \notag
\\
4f^{x,z} &=&2\mathrm{Re}[F^{A,A}\bar{F}^{B,A}-\bar{F}^{B,B}F^{A,B}]  \notag
\\
4f^{y,z} &=&-2\mathrm{Im}[F^{A,A}\bar{F}^{B,A}-\bar{F}^{B,B}F^{A,B}].
\end{eqnarray}%
These results are obtained after replacing the pseudospinors by pseudospin
direction cosines using 
\begin{eqnarray}
2\bar{z}^{A}z^{A} &=&1+n^{z}  \notag \\
2\bar{z}^{B}z^{B} &=&1-n^{z}  \notag \\
2\bar{z}^{A}z^{B} &=&n^{x}+in^{y}
\end{eqnarray}%
Note that because the exchange energy is real, all the interaction form
factors are real. The diagonal interaction form factors are even functions
of $\mathbf{q}$, while the off-diagonal form factors have both even and odd
contributions. The interaction form factors capture the influence of the
shape of the pseudospin state dependent cyclotron orbits on exchange
energies. In the limit $\mathbf{q}\rightarrow 0$, orthogonality implies that 
$F^{A,B}\rightarrow \delta _{A,B}$. If follows that $f^{\alpha ,\beta }(%
\mathbf{q}=0)=\delta _{\alpha ,\beta }/2$.

The exchange energy expression can be written in an alternate form by
Fourier transforming along the direction perpendicular to the guiding center
orbitals, defining 
\begin{equation}
n_{p}^{\alpha }=\frac{1}{N_{\phi }}\,\sum_{X}\,\exp (-ipX)\,n_{X}^{\alpha }
\end{equation}%
where $N_{\phi }=2\pi \ell ^{2}/A$ is the number of guiding center orbitals
in a Landau level. The exchange energy then takes a form from which
spin-wave dispersions can be simply read off: 
\begin{equation}
E_{x}=-\frac{1}{2}\sum_{p}{\bar{n}}_{p}^{\alpha }\,J_{p}^{\alpha ,\beta
}\,n_{p}^{\beta }
\end{equation}%
with the momentum space exchange integral given by 
\begin{equation}
J_{p}^{\alpha ,\beta }=\int \frac{d^{2}\mathbf{q}}{(2\pi )^{2}}\,\exp
(-ipq_{y}\ell ^{2})\,\exp (-q^{2}\ell ^{2}/2)\,f^{\alpha ,\beta }(\mathbf{q}%
)\;V(\mathbf{q}).  \label{jalphabeta}
\end{equation}%
Because $f^{\alpha ,\beta }$ is real, $J_{-p}^{\alpha ,\beta }={\bar{J}}%
_{p}^{\alpha ,\beta }$. Similarly, for the diagonal elements of $J$, the
property that the diagonal elements of $f$ are even functions of $\mathbf{q}$
implies that $J_{p}^{\alpha ,\beta }=J_{-p}^{\alpha ,\beta }$. Combining
both properties, we conclude that the diagonal momentum space interactions
are real. The off-diagonal elements, however, can have both even real and
odd imaginary contributions. For $p\ell \gg 1$, the integration over $%
\mathbf{q}$ in Eq. (\ref{jalphabeta}) has contributions from small $\mathbf{q%
}$ only, allowing us to set $\exp (-q^{2}\ell ^{2}/2)\,f^{\alpha ,\beta }(%
\mathbf{q})\rightarrow \delta _{\alpha ,\beta }/2$. The integral can then be
recognized as an inverse Fourier transform of the electron-electron
interaction from momentum space back to real space. It follows that, 
\begin{equation}
J_{p\rightarrow \infty }^{\alpha ,\beta }=\frac{\delta _{\alpha ,\beta }e^{2}%
}{2\epsilon |p|\ell ^{2}}.
\end{equation}%
This is the familiar peculiarity of quantum Hall systems in which exchange
interactions at large momenta are most simply understood\cite{KallinHalperin}
after a particle hole transformation which converts them into Hartree
interactions between electrons and holes. In strong fields the particle and
hole in a magnetoexciton with momentum $p$ are separated in real space by $%
p\ell ^{2}$. For very large $p$, the separation between particle and hole is
larger than the cyclotron orbit sizes and the interaction is approximately
given by the interaction between point charges. For smaller values of $p$
the size and shape of the cyclotron orbit, represented in momentum space by $%
\exp (-q^{2}\ell ^{2}/2)\,f^{\alpha ,\beta }(\mathbf{q})$ becomes important.

Imaginary off-diagonal contributions to $J_{p}^{\alpha ,\beta }$ are
responsible for DM-like interactions in quantum Hall ferromagnets. In the
example discussed in this paper for instance, the $A$ orbital has orbital
Landau level index $n=0$, while the $B$ orbital has the same spin and $n=1$.
It follows that $F^{A,A}=L_{0}(q^{2}\ell ^{2}/2)$ and $F^{B,B}=L_{1}(q^{2}%
\ell ^{2}/2)$ , where $L_{n}$ is a Laguerre polynomial, and that $%
F^{B,A}=iq\ell \exp (i\theta _{\mathbf{q}})/\sqrt{2}$ where $\theta _{%
\mathbf{q}}$ is the orientation angle of $\mathbf{q}$. It follows that $%
f^{x,y}$ vanishes. The generalized random-phase-approximation that we use in
the main text for collective mode calculations is equivalent to linearized
pseudospin-wave theory. For fluctuations around a ground state with
pseudospin orientation polar angle $\theta _{B}$ (as discussed in the main
text), the $J_{p}^{x,z}$ and $J_{p}^{y,z}$ pseudospin interactions give rise
to DM-like interactions whose strength is proportional to $\sin (\theta
_{B}) $. The DM interactions can be viewed in particle-hole language as a
consequence of the property that the interaction between an electron with
pseudospin in the $\hat{x}-\hat{y}$ plane at $X$ and a hole with pseudospin
in the $\hat{z}$ direction at $X^{\prime }$ is not invariant under the
interchange of $X$ and $X^{\prime }$. This property reflects the pseudospin
dependence of dipole and other higher order multipoles in the cyclotron
orbit cloud of an electron with a particular pseudospin.

The analysis presented in this section, can be applied to any quantum Hall
ferromagnet. One elementary example is the case where the pseudospin
orbitals share Landau level index $n$ and have opposite spins. In this case 
\begin{equation}
f^{\alpha,\beta} \to \frac{\delta_{\alpha,\beta} (L_n(q^2\ell^2/2))}{2} .
\end{equation}
The pseudospin model has isotropic Heisenberg interactions and the
pseudospin-wave excitation energy expression $\omega_{p} = 2 (J_{p=0}-J_{p})$%
, follows immediately from an expansion of the exchange energy function to
quadratic order.

\end{document}